\documentclass[preprint,3p,times,twocolumn]{elsarticle}

\usepackage{amssymb,amsmath,graphicx,float,booktabs}
\usepackage[colorlinks,linkcolor=red,anchorcolor=blue,citecolor=green]{hyperref}
\usepackage{overpic,subfigure}

\usepackage{amssymb,amsmath,graphicx,float,booktabs}
\usepackage{overpic,subfigure}
\usepackage{color}
\usepackage{multirow}
\usepackage{algorithm}
\usepackage{algpseudocode}

\usepackage{cancel}
\usepackage{array}

\newcommand{\mcenter}[1]{\raisebox{-0.5\height}{#1}}
\newcommand{\moffset}[1]{\raisebox{-1.5\height}{#1}}
\newcommand{\moffsetmore}[1]{\raisebox{-20pt}{#1}}
\newcommand{\moffsetup}[1]{\raisebox{30pt}{#1}}

\journal{Computer-Aided Design}

\title{Anti-aliasing for fused filament deposition}

\author[address1]{Hai-Chuan Song}
\author[address1]{Nicolas Ray}
\author[address2]{Dmitry Sokolov}
\author[address1,address2]{Sylvain Lefebvre}

\address[address1]{INRIA, France}
\address[address2]{Universit\'{e} de Lorraine, France}

\begin{document}

\begin{frontmatter}

\begin{abstract}
Layered manufacturing inherently suffers from staircase defects along surfaces that are gently slopped with respect to the build direction.
Reducing the slice thickness improves the situation but never resolves it completely as flat layers remain a poor approximation of the true surface in these regions. In addition, reducing the slice thickness largely increases the print time.

In this work we focus on a simple yet effective technique to improve the print accuracy for layered manufacturing by filament deposition. Our method works with standard three-axis 3D filament printers (e.g. the typical, widely available 3D printers), using standard extrusion nozzles. It better reproduces the geometry of sloped surfaces \textit{without} increasing the print time.

Our key idea is to perform a local \textit{anti-aliasing}, working at a sub-layer accuracy to produce slightly curved deposition paths and reduce approximation errors. This is inspired by Computer Graphics anti-aliasing techniques which consider sub-pixel precision to treat aliasing effects. We show that the necessary deviation in height compared to standard slicing is bounded by half the layer thickness. Therefore, the height changes remain small and plastic deposition remains reliable. We further split and order paths to minimize defects due to the extruder nozzle shape, avoiding any change to the existing hardware. We apply and analyze our approach on 3D printed examples, showing that our technique greatly improves surface accuracy and silhouette quality while keeping the print time nearly identical.

\end{abstract}

\begin{keyword}
Fused filament fabrication, staircase, anti-aliasing, sub-layer, curved layers.
\end{keyword}

\end{frontmatter}

\section{Introduction}

\begin{figure*}\centering{
		\begin{center}
			\subfigure[]{
				\begin{overpic}[width=0.18\linewidth]{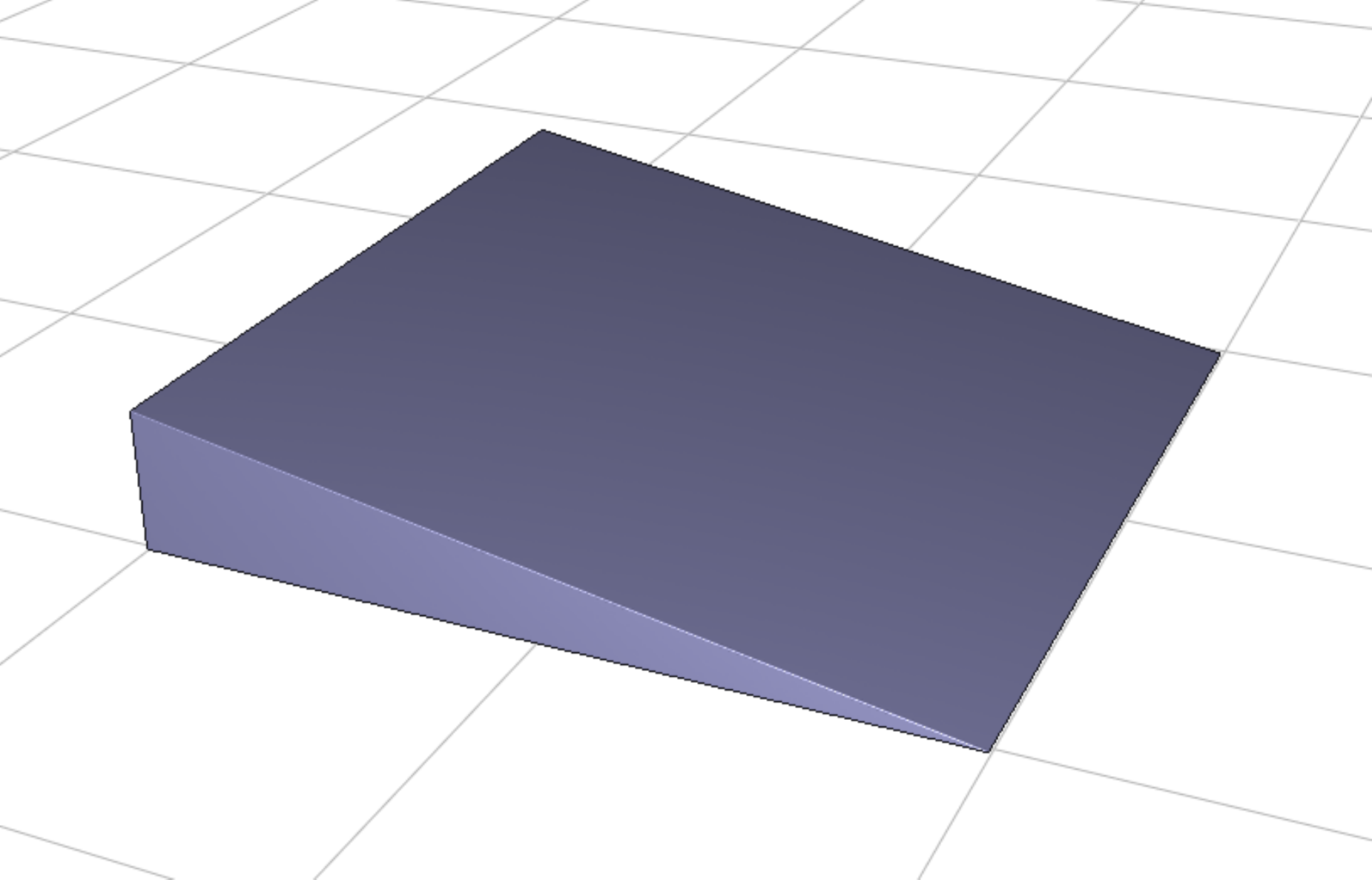}
				\end{overpic}
			}
			\subfigure[]{
				\begin{overpic}[width=0.18\linewidth]{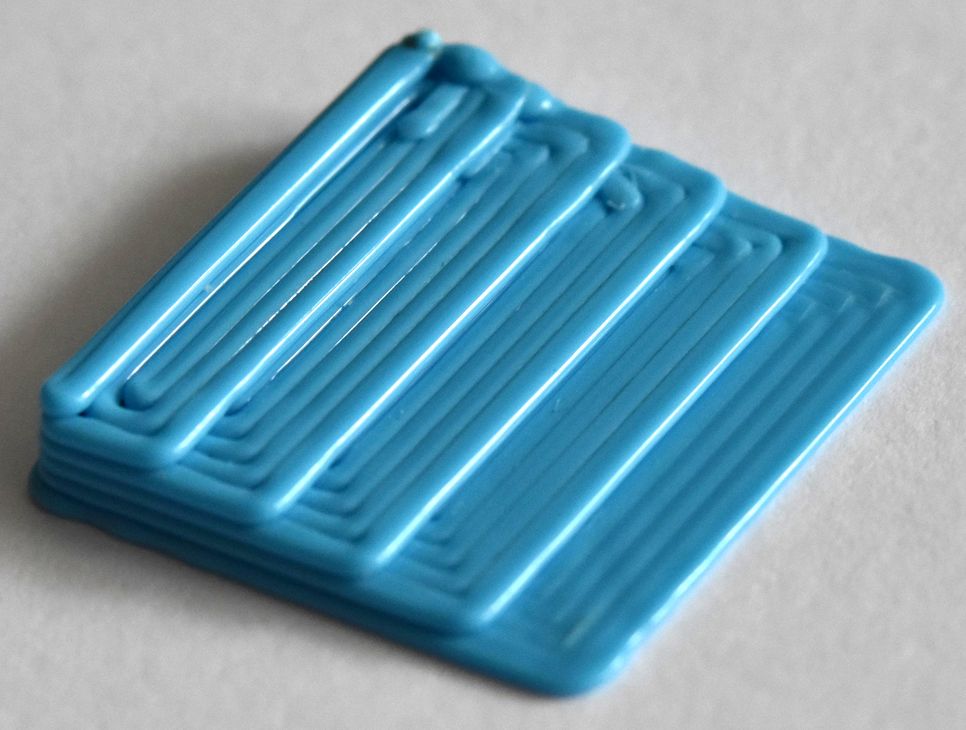}
				\end{overpic}
			}
			\subfigure[]{
				\begin{overpic}[width=0.18\linewidth]{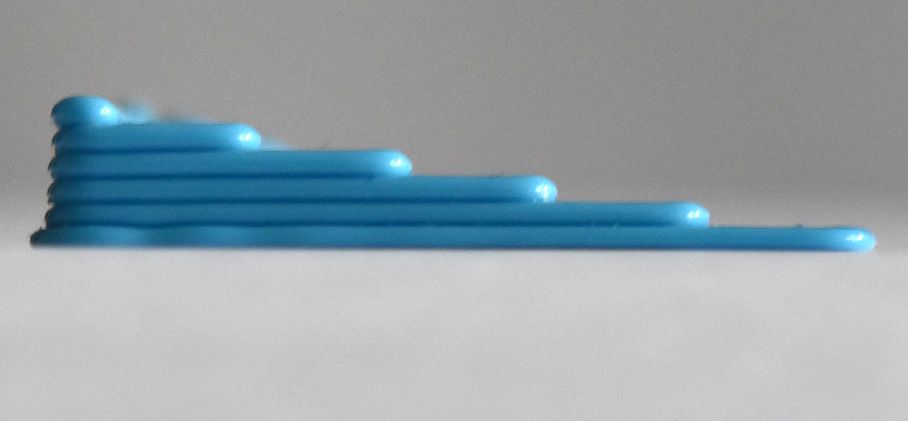}
				\end{overpic}
			}
			\subfigure[]{
				\begin{overpic}[width=0.18\linewidth]{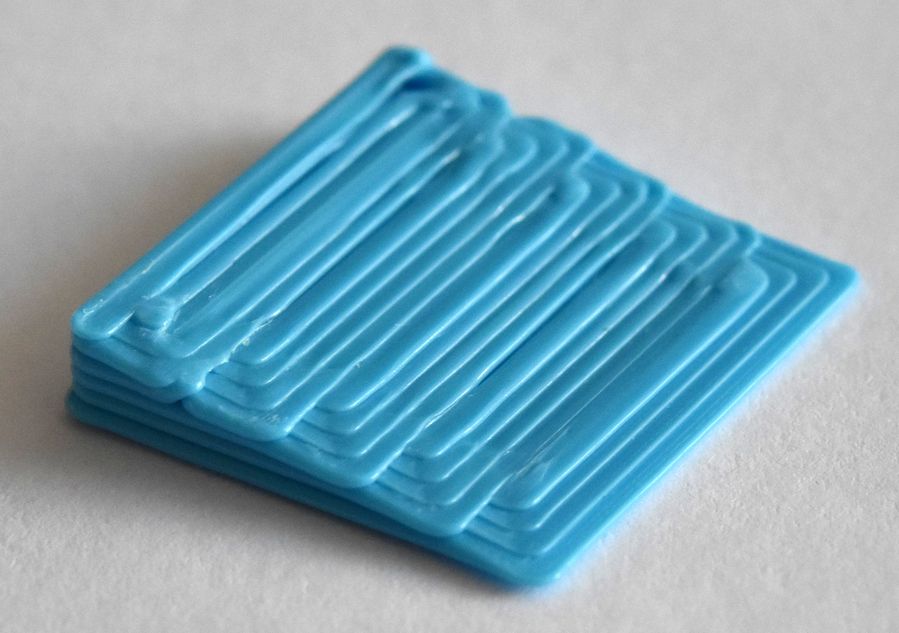}
				\end{overpic}
			}
			\subfigure[]{
				\begin{overpic}[width=0.18\linewidth]{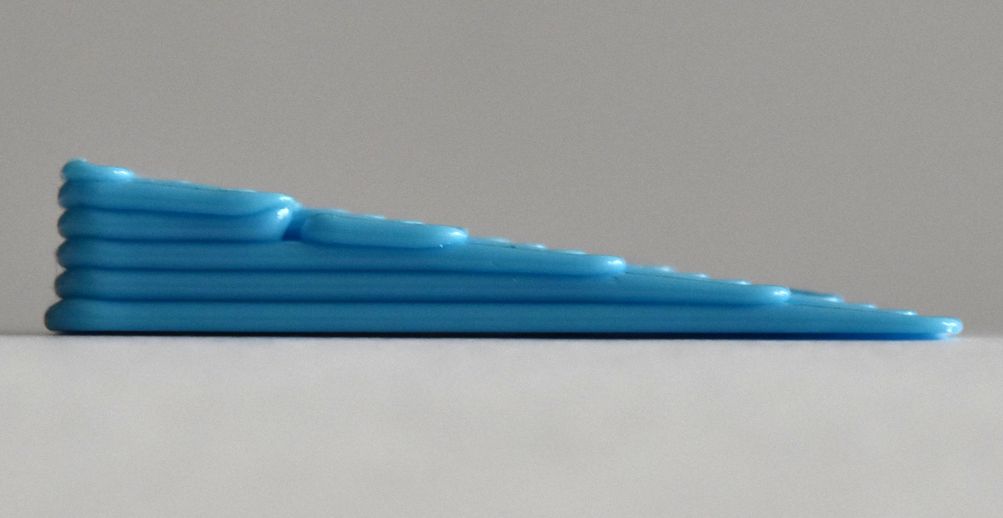}
				\end{overpic}
			}
			\caption{Printing a wedge model clearly reveals the staircase defects that plague 3D printing. (a) Input 3D model ; the bottom edge length is $20$ mm and the angle of the incline plane is $10^\circ$. (b) Global view and (c) side view (c) of a standard, flat layer printed result. (d) Global view and (e) side view of our anti-aliased printed result, revealing the improvement in surface accuracy and silhouette smoothness.}
			\label{fig:teaser}
		\end{center}
}\end{figure*}

Additive manufacturing (AM) technologies produce objects layer-by-layer. The final physical object is thus made of flat slabs of materials stacked on top of each others.
As a consequence, only horizontal and vertical planar surfaces can be closely matched. All other regions suffer from approximation errors, which are often referred to as the \textit{staircase effect}. 
This issue is illustrated in Figure~\ref{fig:teaser} (b) and (c), where a wedge model is sliced into flat layers (layer thickness 0.6 mm) and 3D printed using a filament printer.

\begin{figure}[b]\centering
    \begin{center}
      \begin{overpic}[width=0.25\textwidth]{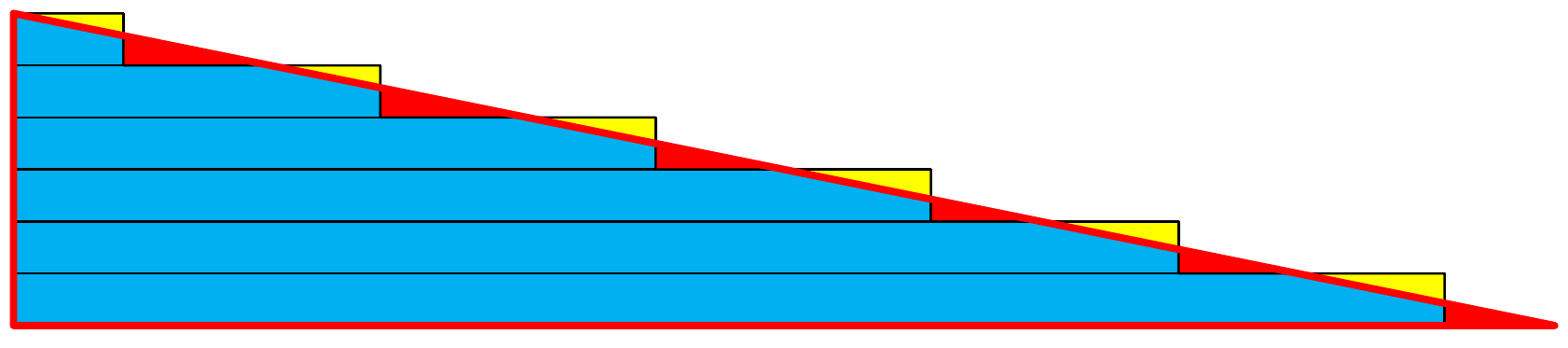}
      \end{overpic}
\caption{Approximation errors of flat layer slicing in the tease: red lines are the input 3D model surfaces; rectangles are printed layers; volumes removed from the input model are colored red; volumes added to the input model are colored yellow.}\label{fig:error_volumes}
    \end{center}
\end{figure}

Such staircase defects lead to a deviation of the printed result from the input model, where necessary volumes in the input model may be removed in the printed result (red regions in Figure \ref{fig:error_volumes}), while unnecessary volumes that do not exist in the input model may appear in the printed result (yellow regions in Figure \ref{fig:error_volumes}). In addition, the final result lacks the smooth geometry of the virtual input model.

A typical approach to minimize staircase defects is to print with thinner layers, which increases the accuracy of the surface approximation. This, however, is achieved at the expense of a large increase in print time (roughly, print time doubles for each reduction of layer thickness by half). The minimal thickness is also limited by material properties -- for filament printers the process becomes challenging under $50\mu m$, even though results down to tens of microns have been demonstrated by experts using precisely calibrated hardware.

Many approaches have been proposed to improve this situation, in particular by using different layer thicknesses in different parts of the model (see Section~\ref{sec:prevwork}). However these approaches still rely on flat layers: the staircase effect is reduced but not removed, and the print time is increased.


We observe that the staircase defect bears similarity with \textit{aliasing} in Computer Graphics, in particular when rasterizing continuous geometry on-screen produces a similar effect. In Computer Graphics, aliasing is often fought by considering information at a sub-pixel scale. We propose to follow a similar idea, and consider sub-\textit{layer} information to produce toolpaths that are subtly curved up and down to follow the surface. Our technique greatly improves surface quality while keeping the print time similar to uniform slicing (see Figure~\ref{fig:teaser} (d) and (e)).

\noindent Our contributions are:
\begin{itemize}
\item The use of sub-layer information to produce (slightly) curved paths that better reproduce the input geometry.
\item A slicing strategy that produces non--flat layers remaining within feasible thickness bounds, never exceeding a minimal thickness threshold.
\item A path splitting and ordering approach that minimizes nozzle interference between neighboring paths of different heights during deposition.
\end{itemize}
\noindent Printed examples reveal a significant increase in the accuracy of the printed surfaces, with print times that remain nearly identical to the initial uniform slicing.

\section{Previous work}
\label{sec:prevwork}

A wide range of approaches have been proposed to minimize the staircase defects. Some apply to different technologies, while others exploit specificities of a particular process. Our work belongs to this second category as our technique focuses on fused filament fabrication: it exploits the possibility to move the extruder up and down during  deposition. 

\paragraph*{Orientation}
Part orientation plays a significant role in determining the impact of the
staircase effect. For a review of this topic, please refer to~\cite{Taufik:2013:ROB}.
Recent works regarding part orientation have focused on using perceptual models to
ensure perceptually important regions are best reproduced~\cite{Zhang:2015:PMO}.

In this work, we assume the object has already been optimally
oriented according to application specific criteria, and focus on precise reproduction given this orientation.

\paragraph*{Adaptive slicing}
Most technologies afford for varying layer thicknesses. Rather than printing with the same uniform thickness throughout the object height, several methods propose to \textit{adapt} the layer thickness to the object geometry~\cite{Dolenc:1994:SPF}.
There are essentially three techniques to decide the layer thickness at each height: subdividing into thinner slices from the coarsest uniform slicing \cite{Sabourin:1996:ASU,Kulkarni:1996:AAS,Hope:1997:ASW}, merging into thicker slices starting from the thinnest uniform slicing \cite{Hayasi:2013:ANA}, or formulating a global optimization problem \cite{Wang:2015:SPS}. Boschetto et al. \cite{Boschetto:2015:TMO} perform an inverse adaptation, deforming the input mesh such that the error after slicing is minimized.
Different error metrics have been proposed to evaluate the error, e.g. the cusp-height~\cite{Dolenc:1994:SPF}, the difference between successive slices~\cite{Zhao:2000}, or the volume error between the original and sliced geometries~\cite{tata1998efficient}.

Our approach is compatible with adaptive slicing (see Section~\ref{sec:antilayers}), and will further improve its results.

\paragraph*{Locally adaptive slicing}

A drawback of adaptive slicing is that it can only adapt to geometric changes along the built direction: the thickness remains constant within the layer. To address this problem several approaches first perform an object decomposition, and then independently slice different regions of the object. The object can be either printed as a single part with careful ordering of the toolpaths~\cite{Sabourin:1997:AEF,Tyberg:1998:LAS,Mani:1999:RBA,Wang:2015:SPS}, or can be printed in several pieces that are later manually assembled~\cite{hildebrand2013orthogonal,Hu:2014:APS,Wang:2016:ISQ}.

Our approach performs local thickness adaptation of the layers through small z-motions (sub-layer precision), avoiding a subdivision of the geometry into distinct parts.

\paragraph*{Curved slicing}
Several approaches exploit the ability of filament printers to move freely along the z-axis during deposition.
Chakraborty et al.~\cite{Chakraborty:2008:EPG} proposed a curved layer deposition, with the objective of strengthening shell-like parts by aligning the toolpaths with the surface. The mechanical properties of the parts are further discussed in~\cite{Singamneni:2012:MAE}, and methods combining flat and curved layers are discussed in~\cite{Huang:2012:ASA,Allen:2015:AED}. In these works the curvature of the paths is large as they flow along the surface (e.g. several millimeters up and down). While this works very well on specific cases, in a general setting this requires specifically designed hardware to avoid interferences between the nozzle and the printed part as discussed
\footnote{Quote from~\cite{Chakraborty:2008:EPG} : ``the periphery of the deposition head could interfere with the part being built up (..) deposition head shape would have to be modified (..) to avoid such phenomenon";
Quote from~\cite{Allen:2015:AED}: ``A solution to track distortion is to maintain normality between the extruder axis and (..) surface (..). Although not feasible with the current system in use, robots with sufficient degrees of freedom (..)"
} in~\cite{Chakraborty:2008:EPG,Allen:2015:AED} and Figure~\ref{fig:erase_path}.

Our approach only has to perform small sub-layer adjustments, requiring no change to the printer as the layer interference remains manageable (but still requires careful treatment). However, our work has slightly different objectives since we seek to increase the accuracy of the surfaces, while the aforementioned works attempt to align the filament deposition with the surface curvature to produce stronger parts.

In a more extreme demonstration of the opportunities for non-flat deposition, Mueller et al.~\cite{Mueller:2014:WDP} propose to print wire-mesh structures, exploiting the fact that extruded filament hardens quickly to print truss-like structures in mid-air.

\paragraph*{Process based surface quality improvements}

While mechanical means can be used to improve surface quality after fabrication (e.g. abrasion~\cite{Williams:1998:AFF} or machining~\cite{Pandey:2006:VHF}), researchers have proposed a number of software techniques that exploit specificities of a process to reduce the staircase effect. Most notably, in the context of stereo-lithography (SLA) Pan et al.~\cite{Pan:2012:SSF,Pan:2015:SSF} exploit the formation of a meniscus when an object moves out of the resin tank. The meniscus is cured to fill the creases between two layers. Repeating this process produces smooth, accurate surfaces.
Park et al.~\cite{Park:2011:DMF} show how dithering can cure resin partially and produce slanted surfaces along a layer.

Our work also exploits sub-layer precision, relying on properties specific to fused filament fabrication.

\subsection{Printing with varying layer thickness}
\label{sec:varyingprint}

\begin{figure}\centering
	\begin{center}
		\begin{overpic}[width=0.4\textwidth]{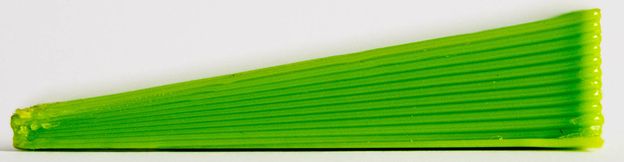}
		\end{overpic}
		\caption{A slope generated by varying layer thicknesses across 12 layers. Each layer thickness varies from 0.07 mm to 0.38 mm uniformly from left to right. The nozzle width is 0.4 mm.}
		\label{fig:varying_path_thickness}
	\end{center}
\end{figure}

We provide here some background on the technical aspects of printing with varying layer thicknesses.
In printer instructions (G-code) each print path is a sequence of vertices, each having a 3D coordinate $(x,y,z)$ as well as a length of extruded filament $e$. Thus, varying
the layer height involves updating $z$ and $e$ in a synchronized manner such that the width of deposited
plastic remains regular.
Therefore, if the vertex is displaced vertically by $\delta$ the filament extrusion length becomes~$\frac{z+\delta}{z}\cdot{e}$.

We found that reducing the extrusion speed (the length of filament extruded per second) when changing the thickness provides a more regular extrusion. Given an initial speed $F_{ini}$, a minimum speed $F_{min}$, the vertical displacements of two neighboring vertices $\delta_1$, $\delta_2$ and a base layer thickness $h$, we compute the new speed as a linear interpolation $F = {F_{ini}} + \frac{{\left| {{\delta_1} - {\delta_2}} \right|}}{h}\cdot\left( {{F_{min}} - {F_{ini}}} \right)$.
This slows down where there are larger changes in height. We use $F_{ini} = 20$ mm/sec and $F_{min}=13$ mm/sec in all our results. These are standard speeds for good quality prints.
%
Figure~\ref{fig:varying_path_thickness} illustrates a slope generated by varying layer thicknesses. 
	

\begin{figure}[b]\centering
	\begin{center}
			\begin{overpic}[width=0.4\textwidth]{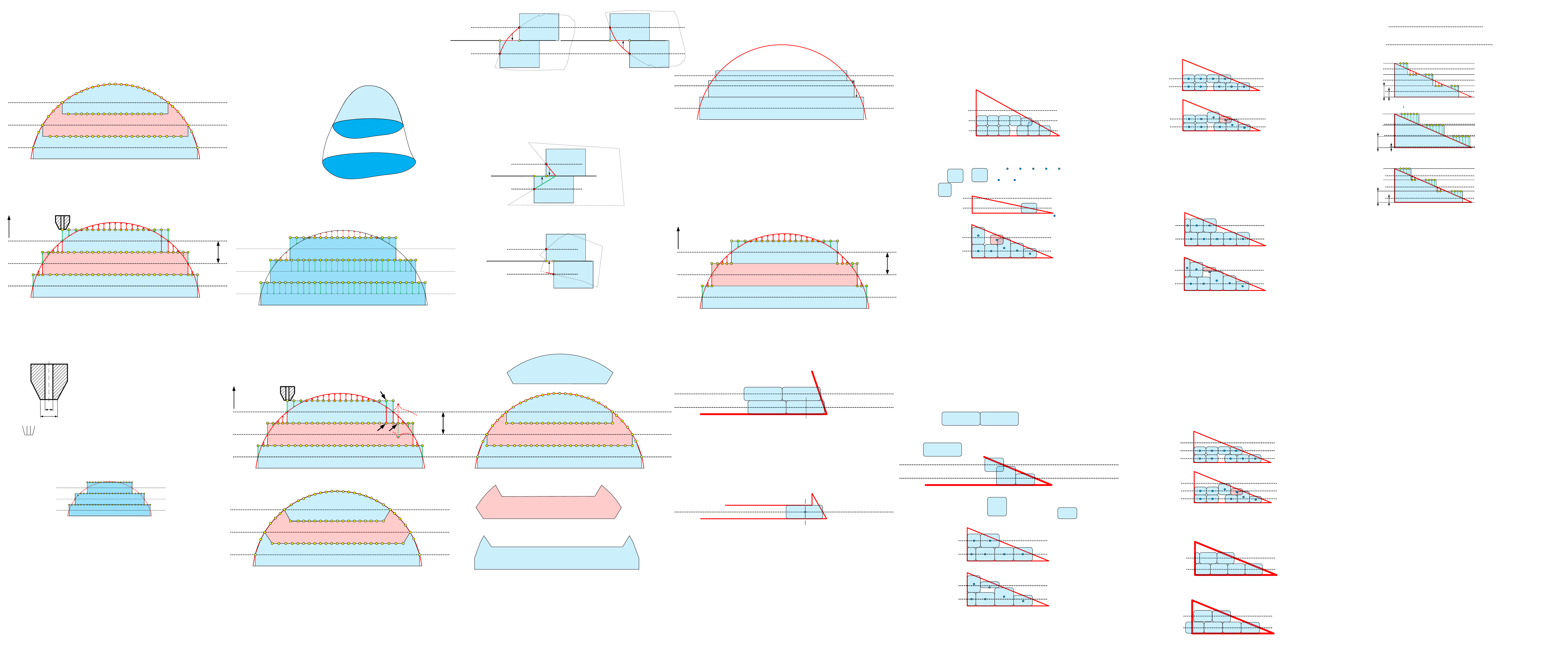}
				\put(91.3,19.5){$h$}
				\put(3,34){$Z$}
			\end{overpic}
		\caption{2D illustration of uniform slicing (side view). The red curve is the input model boundary. The blue and red rectangles are cut view of the layers. Dashed lines are the mid-layer slicing planes. The layers are vertical extrusions of the intersection between the slicing planes and the initial object.
			The yellow circles are points on the top of the layers, that form the physical object boundary. The arrows reveal the distance between the printed surface and the true surface (red: positive; green: negative).
			 }
		\label{fig:flat_slice}
	\end{center}
\end{figure}

\section{Toolpath anti-aliasing}

The key idea of our technique is better explained by considering the shape of each layer. After uniform slicing, all layers are flat. Conceptually, our approach locally deforms the input flat layers to better approximate the surface. This is detailed in Section~\ref{sec:antilayers}, where we also show that these variations do not need to exceed half the layer thickness, up or down.

In practice however, our method operates primarily on the toolpaths produced by a standard uniform slicing process.
Each toolpath is a sequence of vertices, forming a curve along which the extruder deposits plastic.
Our algorithm performs the following operations on each layer:
\begin{itemize}
\item For each toolpath vertex, the local \textit{vertical} distance to the surface of the input model is computed through an efficient ray-tracing procedure.
\item The paths are then deformed (\textit{anti-aliased}) to better capture the sub-layer variations of the surface. This is detailed in Section~\ref{sec:layer2toolpaths}.
\item Finally, the paths are split and ordered to minimize interferences with the extruder nozzle. This is detailed in Section~\ref{sec:interference}.
\end{itemize}
The last step is important to the success of our approach: it minimizes cases where a currently printed path collides with (and erases) a previously printed paths that was raised higher.

\subsection{Layer anti-aliasing}
\label{sec:antilayers}

The input model is sliced into flat layers by intersecting slicing planes with the 3D model. The slicing planes are located at the middle section of each slice, following a standard slicing process.
A side view example is shown in Figure \ref{fig:flat_slice}. The 3D model is represented by the red curve and sliced into three flat layers with the same thickness $h$.
The dashed lines outline the mid-layer slicing planes. The extruder deposits plastic by moving along the top plane of the layers, forming vertical plastic slabs below.


Slicing a curved surface with flat layers produces a staircase defect, clearly visible in Figure \ref{fig:flat_slice} : the tops of the printed layers do not follow the original model boundary. To reduce aliasing, we propose to deform the layers and align their boundaries back with the original surface. This changes the layer shapes and produces corresponding z-motions of the extruder.


\begin{figure}\centering
	\begin{center}
		\subfigure[]{
			\begin{overpic}[width=0.19\textwidth]{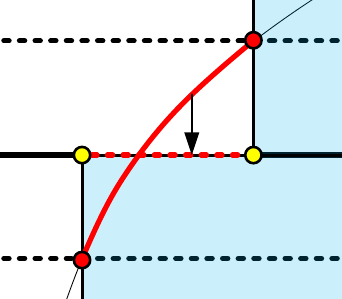}
				\put(53,62){$P$}
				\put(53,34){$P_L$}
                \put(-15,8){$L$}
                \put(-25,72){$L+1$}
			\end{overpic}
		}
		\subfigure[]{
			\begin{overpic}[width=0.185\textwidth]{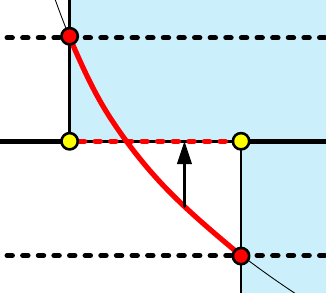}
				\put(52,18){$P$}
				\put(52,49){$P_L$}
			\end{overpic}
		}
		\caption{
			Illustration of the mapping between surface points and layers for the case of respectively top-facing (a) and bottom-facing (b) surfaces.
			The dashed black lines are mid-layer slicing planes. The dashed red segment is a physical boundary of the printed shape, at the top (a) or bottom (b).
			Each surface point $P$ vertically projects to a point $P_L$ on this segment. If the mapping is injective we can snap the points $P_L$ back to the surface.
			The displacement is $\pm \frac{h}{2}$ at most.
			}\label{fig:mapping}
	\end{center}
\end{figure}
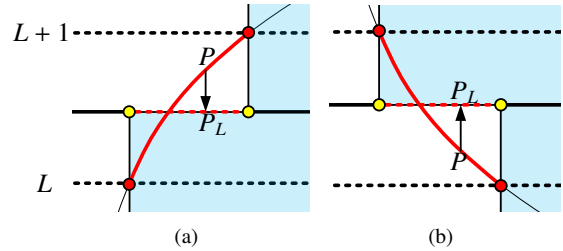

\paragraph*{Thickness bounds}
An important constraint when modifying the layer shapes is to maintain the local thickness within feasible bounds, otherwise plastic extrusion and adhesion fails. For plastic filament deposition, our experiments reveal a minimal thickness of $0.05$ mm and a maximum thickness of about $1.1 w$, with $w$ the nozzle width (we tested with $w=0.4$ mm and $w=0.8$ mm). The lower bound is due to the physical process, but also to errors in the printer assembly and calibration that stop being negligible with small thicknesses.

We deformed, the layers near the boundary may stretch too much and exceed these bounds. A possible approach would be to diffuse the deformation from the boundary layers to the inner layers, producing curved layers throughout the print. This leads back to curved layer slicing (see Section~\ref{sec:prevwork}) which presents many challenges.
Instead, we propose a simpler, yet effective solution that ensures the layer thickness always remains within $[\frac{h}{2},  \frac{3 \cdot h}{2}]$. The new layers properly capture all surface points but for extreme cases (discussed later). This approach is compatible with adaptive slicing.

\paragraph*{Deforming the layers}
After slicing, each layer is a vertical extrusion of the contour at the mid-layer slicing plane (see Figure~\ref{fig:flat_slice}). We independently deform the top and bottom of the layers to better reproduce respectively the top-facing and bottom-facing parts of the print. Even though we describe how to change the bottom shape of the layers, please note that our current implementation does not treat down-facing surfaces.


For the sake of clarity, let us first consider only the top-facing parts of the surface.
We denote $P$ a surface point, and it is uniquely enclosed between the slicing planes of two layers (black dashed lines in Figure~\ref{fig:mapping}). We denote the layers $L$ and $L+1$. We denote $P_L$ the vertical projection of a point $P$ on the top of layer $L$ (resp. bottom of layer $L+1$), as illustrated in Figure~\ref{fig:mapping}. The vertical distance between $P$ and $P_L$ is $\delta_{P_{L}} = z(P) - z(P_L)$ where $z(.)$ returns the z axis coordinate.
Note that we always have $-h/2 \leq \delta_{P_L} \leq h/2$.

Let us assume that a unique point $P$ projects to a same point $P_L$, that is the projection from $P$ to $P_L$ is injective as in Figure~\ref{fig:mapping}. This assumption is generally true since most printed objects have a feature size much larger that the layer thickness. The point $P$ lies above the slicing plane of $L$ but below the slicing plane of $L+1$~; thus $P_L$ belongs to a top part of the printed surface, visible in Figure~\ref{fig:mapping} (a) as the dashed red segment. Indeed, layer $L+1$ will not deposit any material at these locations (all points $P$ project outside the layer contour since they are below the slicing plane). Conversely, all points $P$ project inside the contour of layer $L$, and layer $L$ deposits the material forming the top surface. Thus, for each $P_L$, displacing the top surface of layer $L$ by $\delta_{P_{L}}$ will align the layer and the surface. None of these displacements exceed $|h/2|$ in magnitude.

\begin{figure}[t]\centering
	\begin{center}
		\subfigure[]{
			\begin{overpic}[width=0.17\textwidth]{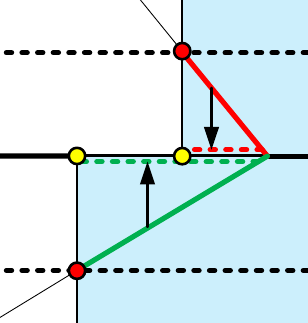}
				\put(43,21){$P$}
				\put(43,54){$P_L$}
				\put(62,75.5){$Q$}
				\put(62,44){$Q_L$}
			\end{overpic}
		}
		\subfigure[]{
			\begin{overpic}[width=0.185\textwidth]{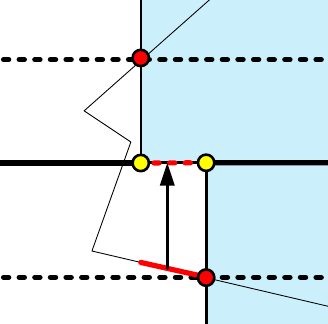}
				\put(47,6){$P$}
				\put(47,52){$P_L$}
			\end{overpic}
		}
		\caption{\textbf{(a)} Displacing both the bottom and top of layers correctly opens a tear between two adjacent layers.
			\textbf{(b)} Surfaces folding more than once within the slice cannot be properly approximated.}\label{fig:mapping2}
	\end{center}
\end{figure}

\begin{figure}[t]\centering
	\begin{center}
		\begin{overpic}[width=0.2\textwidth]{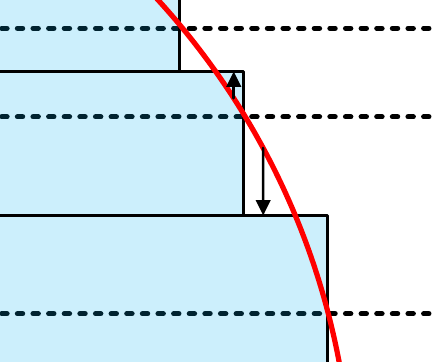}
            \put(47,57){$P$}
			\put(50,69){$P_L$}
			\put(62,49){$Q$}
			\put(58,27){$Q_L$}
		\end{overpic}
		\caption{Mapping between surface points $P$ and $Q$ and layer tops/bottoms in adaptive slicing (respectively $P_L$ and $Q_L$).
			The magnitude of the displacements remains bounded by half the layer thickness, but the actual value depends on which layer top/bottom is displaced.}
		\label{fig:adaptive}
	\end{center}
\end{figure}
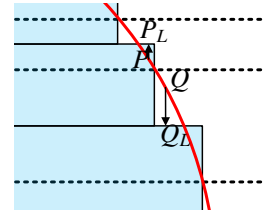

This same observation applies for the down-facing part of the surface, this time displacing the bottom surface of layer $L+1$, as illustrated in Figure~\ref{fig:mapping} (b). Interestingly, this definition correctly opens a 'tear' between layers where the shape exhibits a vertical concavity, as illustrated in Figure~\ref{fig:mapping2} (a). This approach is compatible with adaptive slicing, as illustrated in Figure~\ref{fig:adaptive}.

With this definition, as long as the mapping between surface points $P$ and layer tops/bottoms $P_L$ remains injective, each surface point gets covered by a displaced layer top/bottom. Figure~\ref{fig:mapping2} (b) shows a counter example where the surface folds more that once within the slice. In such a case, the displaced layers cannot capture the surface of the input model.. Note that this limitation is intrinsic to mid-layer slicing, which only produces optimal slices when the object shape is monotonous within the slice~\cite{hildebrand2013orthogonal}.

We next discuss our practical implementation of this approach, starting from the toolpaths produced by a standard slicer.


\subsection{From layers to toolpaths}
\label{sec:layer2toolpaths}

Given the definition of anti-aliased layers from Section~\ref{sec:antilayers}, we could proceed with generating toolpaths within these new layers.
Instead, we make our approach compatible with existing slicers by deforming the \textit{toolpaths} produced by a standard slicer.

When input a closed, two-manifold non self-intersecting geometry, planar slicing produces 2D closed contours within each layer.
From each contour the slicer extracts toolpaths for the external perimeters, inner contours, as well as infill paths (see \cite{dinh.15.sigcourse} for more details).

We propose to work directly from these toolpaths. Our algorithm starts by re-sampling the paths, inserting vertices in path segments until their length
is below the nozzle diameter.
We then trace a vertical ray starting from each path vertex towards the model surface above and below. We keep the closest surface point $P$ which gives the distance $\delta_{P_L}$.
When $P$ belongs to the top-facing part of the surface and the distance is within the range $[-h/2,h/2]$ we displace the vertex by $\delta_{P_L}$, otherwise we leave it untouched.
This effectively displaces the toolpaths towards the top-facing surfaces.

Clearly, this approach can only capture surface details which are bigger than the width of a toolpath (typically $w$): the toolpath vertices are a subset of the
points $P_L$ from Section~\ref{sec:antilayers} and cannot be used to reproduce an arbitrarily detailed relief. We will see in Section~\ref{sec:results} that this
results in some limitations. However this improves all surfaces which smoothly vary compared to $w$ -- the vast majority of cases, and the cases which produce the worst
cases when sliced with flat layers.

Our current implementation does not treat down-facing surfaces. However, the following procedure could be used. Let us assume a dense support from below, sliced using our anti-aliasing algorithm. The support is the complement of the bottom-facing surface, and therefore its top-facing surfaces will have been treated properly by our algorithm. Thus, for the surface we do not need to displace the toolpath vertices. Instead, the plastic flow is adjusted such that the layer is locally thinner. After support removal (e.g. chemical dissolve), the bottom surface will be correctly anti-aliased from below.


\subsection{Overlaps and flow adjustment}
\label{sec:flow}

\begin{figure}\centering
	\begin{center}
		\subfigure[]{
			\begin{overpic}[width=0.22\textwidth]{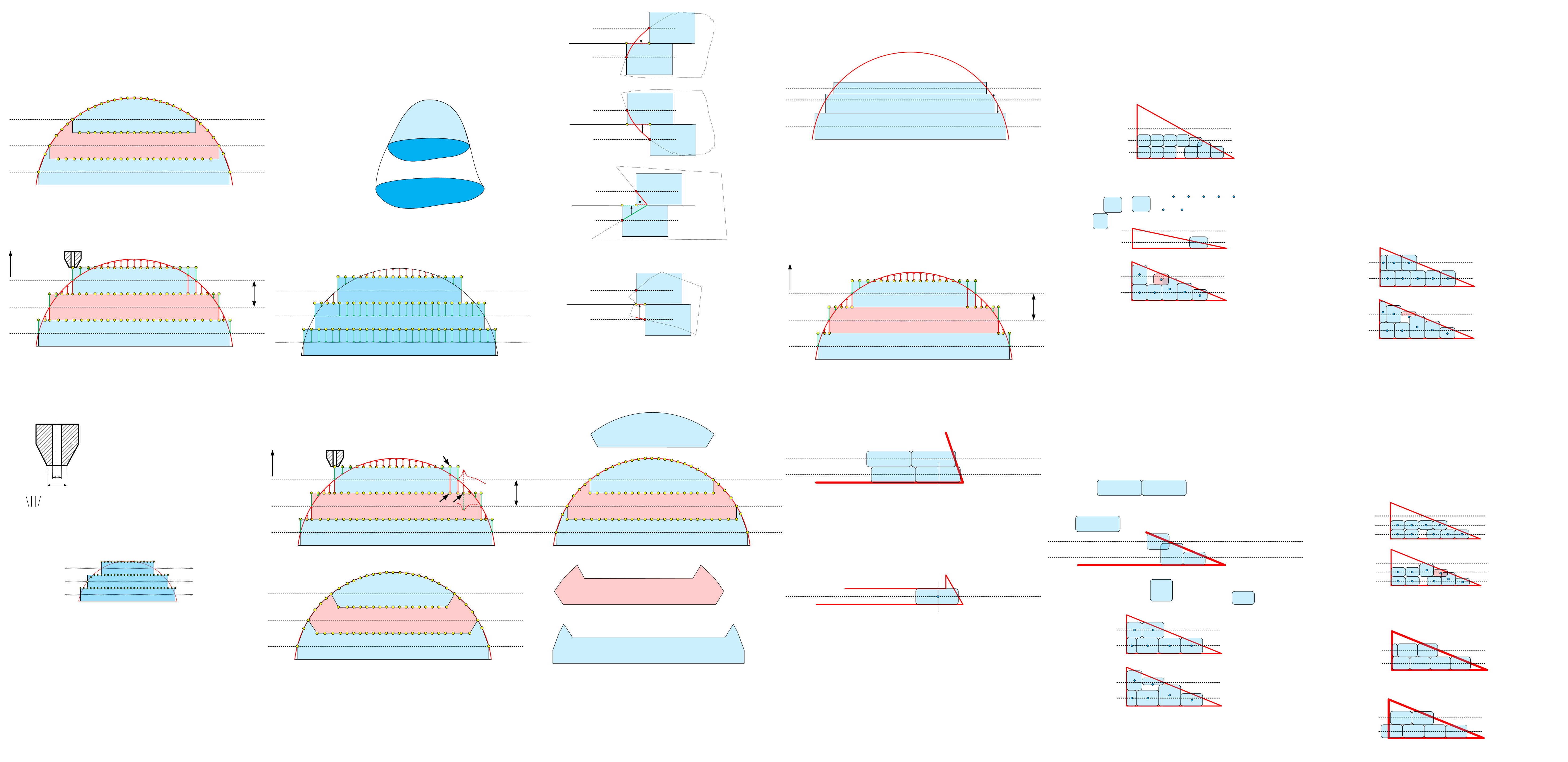}
			\end{overpic}
		}
		\subfigure[]{
			\begin{overpic}[width=0.22\textwidth]{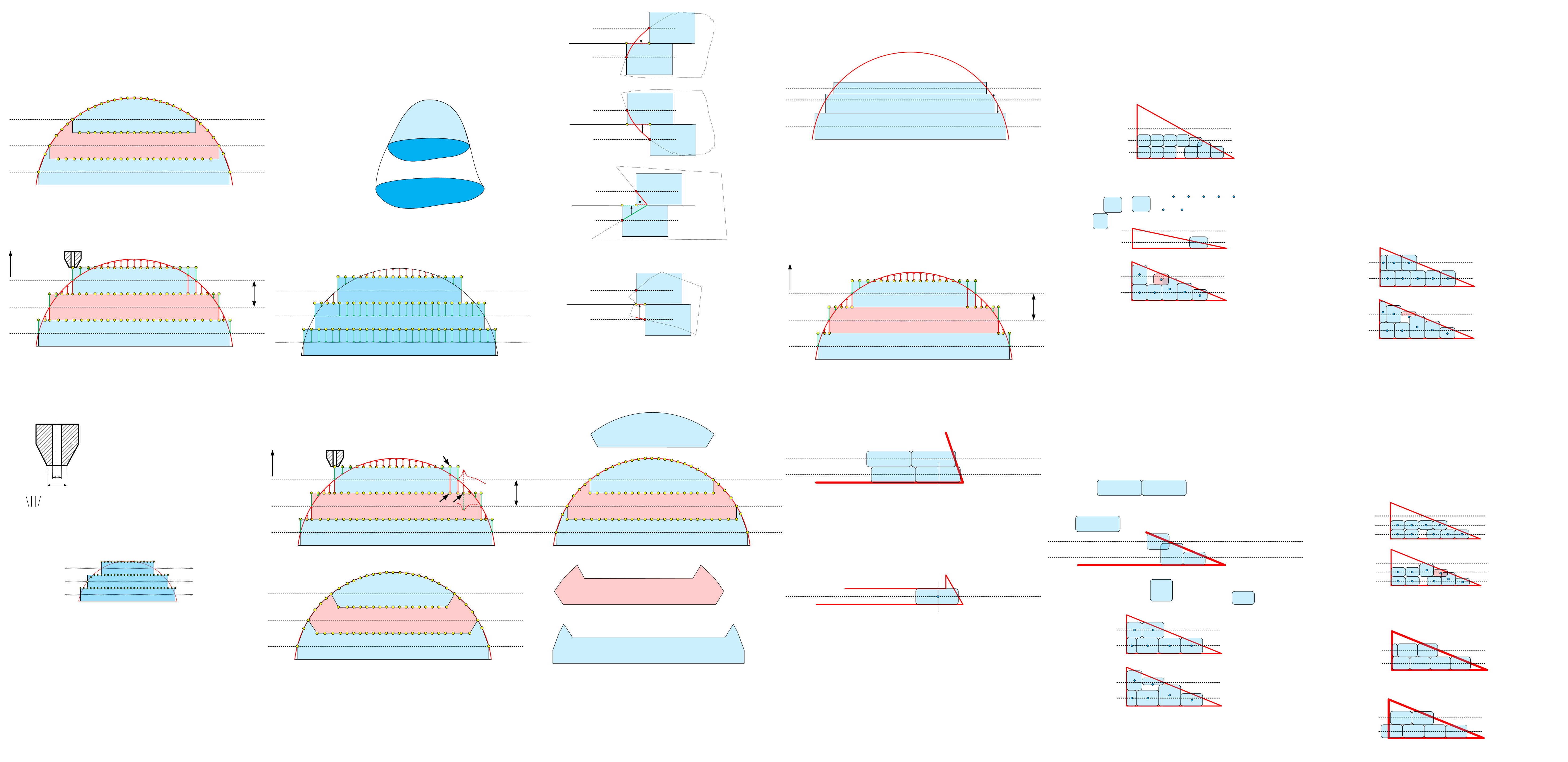}
			\end{overpic}
		}
		\caption{Raising the toolpaths may produce small overlaps where toolpaths do not align across layers. \textbf{(a)} Original uniform slice, viewing the cross section of the deposited plastic tracks. \textbf{(b)} Anti-aliased layer with displaced plastic tracks, revealing the overlap.}
				\label{fig:overlaps}
	\end{center}
\end{figure}

In general the toolpaths do not align across layers. As a consequence raising a toolpath might generate a small overlap with a toolpath located above, as illustrated in Figure~\ref{fig:overlaps}. This is again due to the discretization of the layers by the toolpaths: the deformed layers do not overlap themselves.

To avoid any detrimental effect on the final print we adjust the plastic flow. In particular, we detect the overlap between toolpath segments and reduce the flow by the volume of the intersection.

\subsection{Slicing plane position}
\label{sec:slicingplane}

In Section~\ref{sec:antilayers} we have considered the slicing planes to be in the middle of the layers.
This is the typical setup and ensures that our approach is compatible with the toolpaths
produced by any slicer.
However, we can also change the location of the slicing plane, moving it anywhere within the layer
from $0$ to $h$ (default being $\frac{h}{2}$).

\begin{figure}[t]\centering
	\begin{center}
		\subfigure[]{
			\begin{overpic}[width=0.22\textwidth]{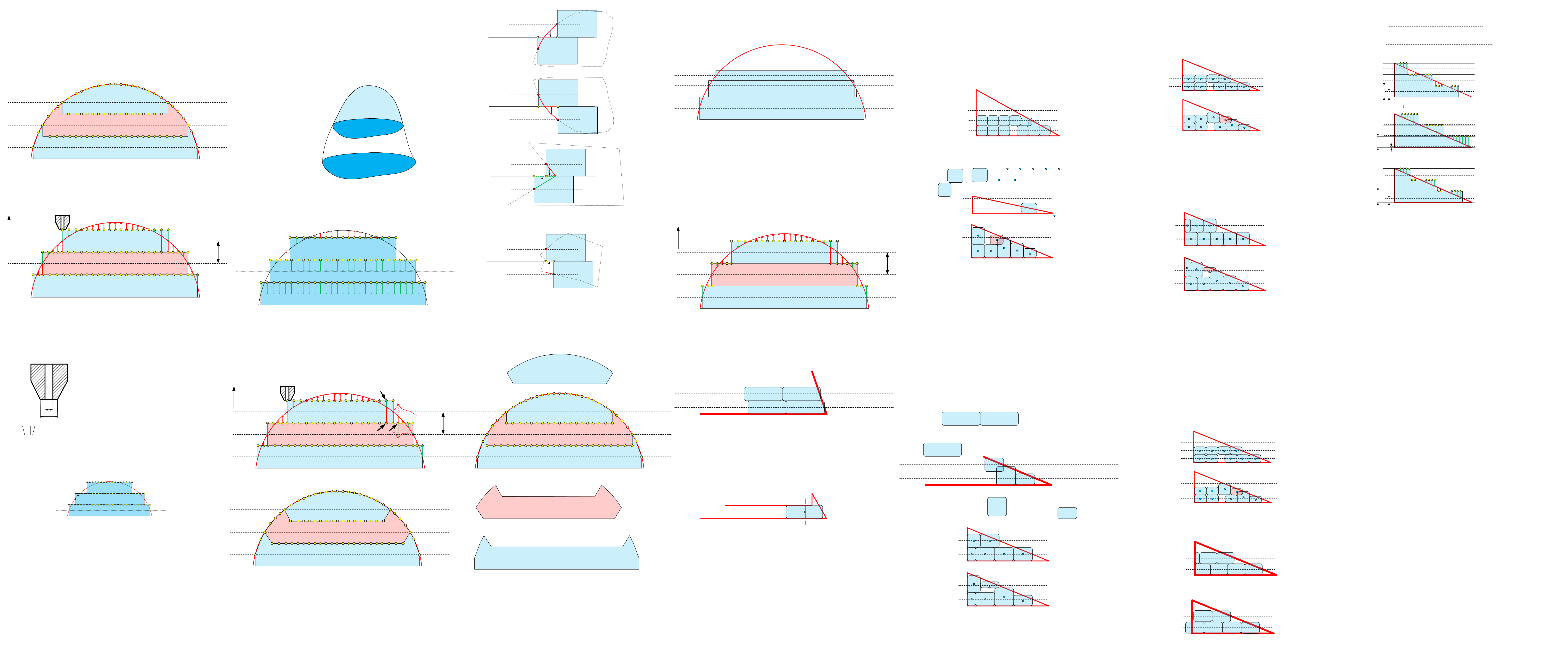}
            \put(2.3,6.4){\tiny{$s=0$}}
            \put(-1.5,8){\tiny$h$}
			\end{overpic}
		}
		\subfigure[]{
			\begin{overpic}[width=0.22\textwidth]{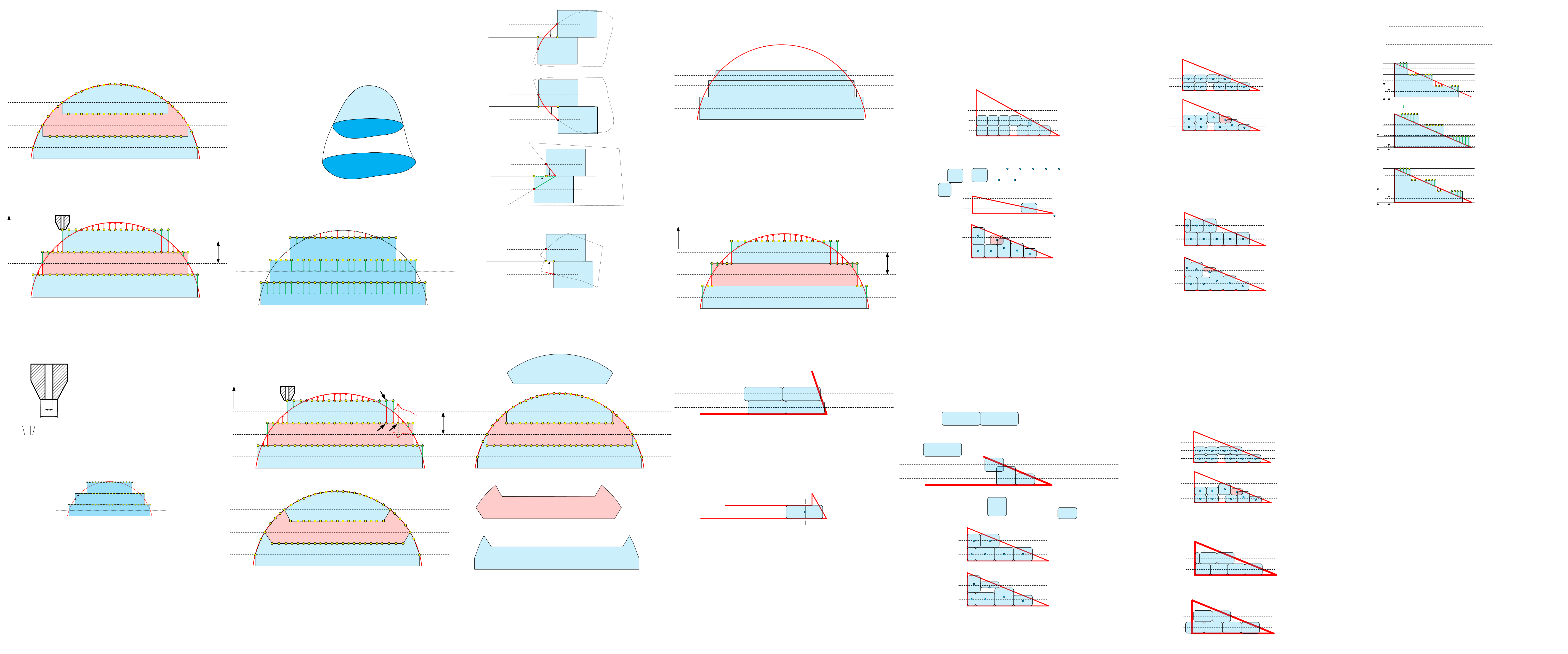}
            \put(9,6){\tiny{$s$}}
            \put(-2,9){\tiny{$h$}}
			\end{overpic}
		}
		\caption{\textbf{Left:} $s=0$ produces only downwards displacements, but exceeds the lower thickness bound. \textbf{Right:} Intermediate values less than $\frac{h}{2}$ exploit thinner slices while reducing overlaps across layers.}
		\label{fig:slicingplane}
	\end{center}
\end{figure}

Let us denote $s \in [0,h]$ the position of the slicing plane. We have seen that for $s=\frac{h}{2}$, layer
displacements are in $[-\frac{h}{2},\frac{h}{2}]$. Figure~\ref{fig:slicingplane} illustrates the displacements that would occur for other values of $s$.

Interestingly, setting $s=0$ only requires downwards displacements, i.e. within $[-h,0]$. This avoids overlaps across layers (see Section~\ref{sec:flow}). However, this also produces layers where the thickness goes \textit{below} the lower feasible bound.

\begin{figure}[t]\centering
	\begin{center}
	\includegraphics[width=1.0\linewidth]{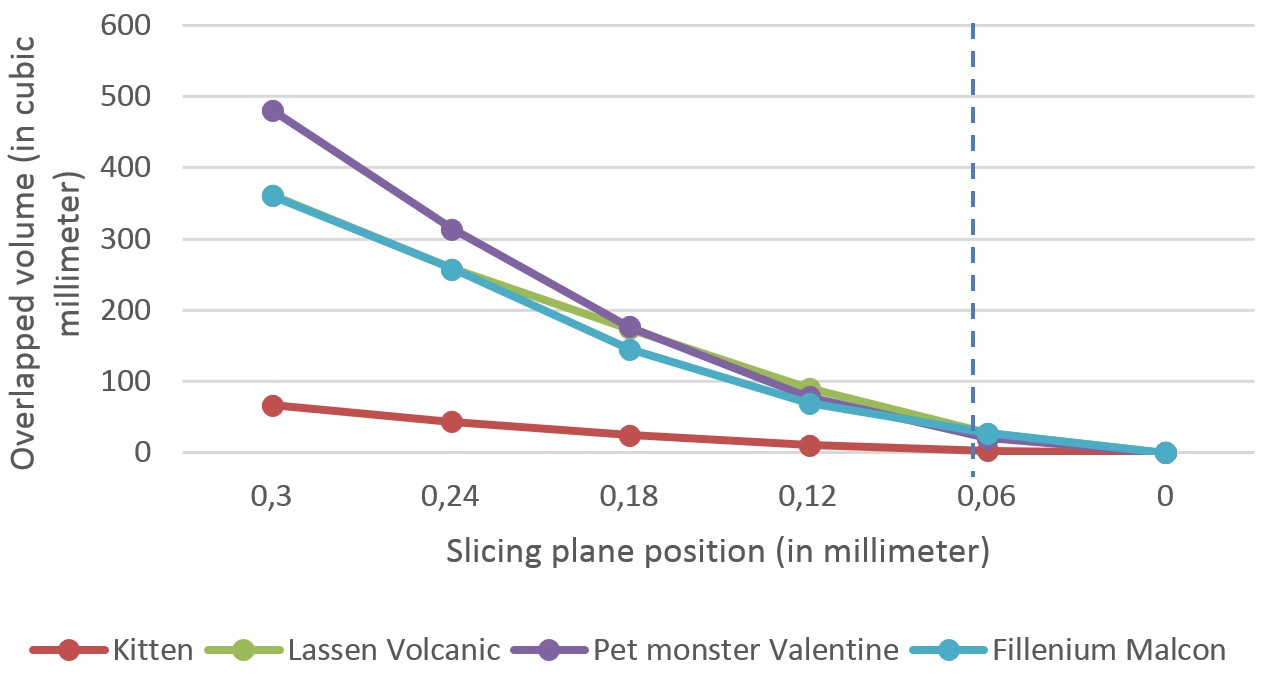}
	\caption{Volume of overlap versus slicing plane position. The dashed line indicates the slicing position with the thinnest possible layers (0.06 mm on our printer). Models are shown in the result section.}
	\label{fig:overlap_vs_position}
	\end{center}
\end{figure}

Setting $s$ to any intermediate value produces displacements within the range $[s-h,s]$. A benefit of not following the standard procedure would be to select $s$ at the smallest value where deposition is reliable, e.g. $0.06$ mm. For our setup, where $w=0.8, h=0.6$ this leads to thicknesses within $[0.06,0.54]$. This fully exploits the capability to produce thin layers while strongly reducing potential overlaps across layers. However, this requires changes to the slicer code to adjust the slicing plane position.
Figure~\ref{fig:overlap_vs_position} reports the potential benefits in terms of overlap reduction for our test models.

We only used the standard setup $s=\frac{h}{2}$ in our results to enforce backward compatibility.

\section{Interference avoidance}
\label{sec:interference}

\begin{figure}\centering
	\begin{center}
		\begin{overpic}[width=0.15\textwidth]{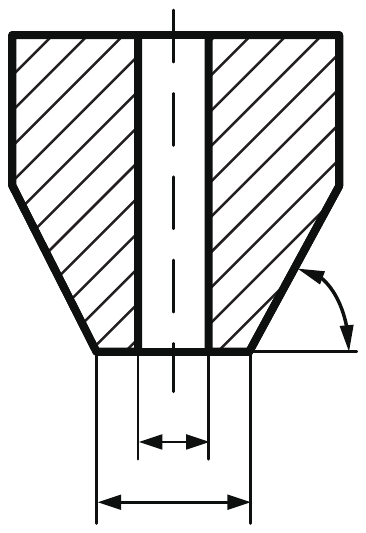}
			\put(30,19){$w$}
			\put(30,7){$\tau$}
            \put(64,43){$\alpha$}
		\end{overpic}
		\caption{Typical nozzle geometry, sectional view. $w$ is the inner diameter, $\tau$ is the outer diameter, $\alpha$ is the nozzle sides inclination.}
	\label{fig:nozzle}
	\end{center}
\end{figure}

During plastic deposition the active nozzle moves alongside previously printed paths. With flat layers this is never a problem as the nozzle cannot collide with previously printed paths: all have the same height.
However, our algorithm produces toolpaths of varying heights within a same layer. Even though, the height variations are small ($\pm h/2$) they are enough to make the nozzle interfere with printed paths.
This is worsen by the nozzle designs which are optimized for printing flat layers and typically have a wide flat end, as illustrated Figure~\ref{fig:nozzle}.
\begin{figure}\centering
	\begin{center}
		\subfigure[]{
			\begin{overpic}[width=0.3\textwidth]{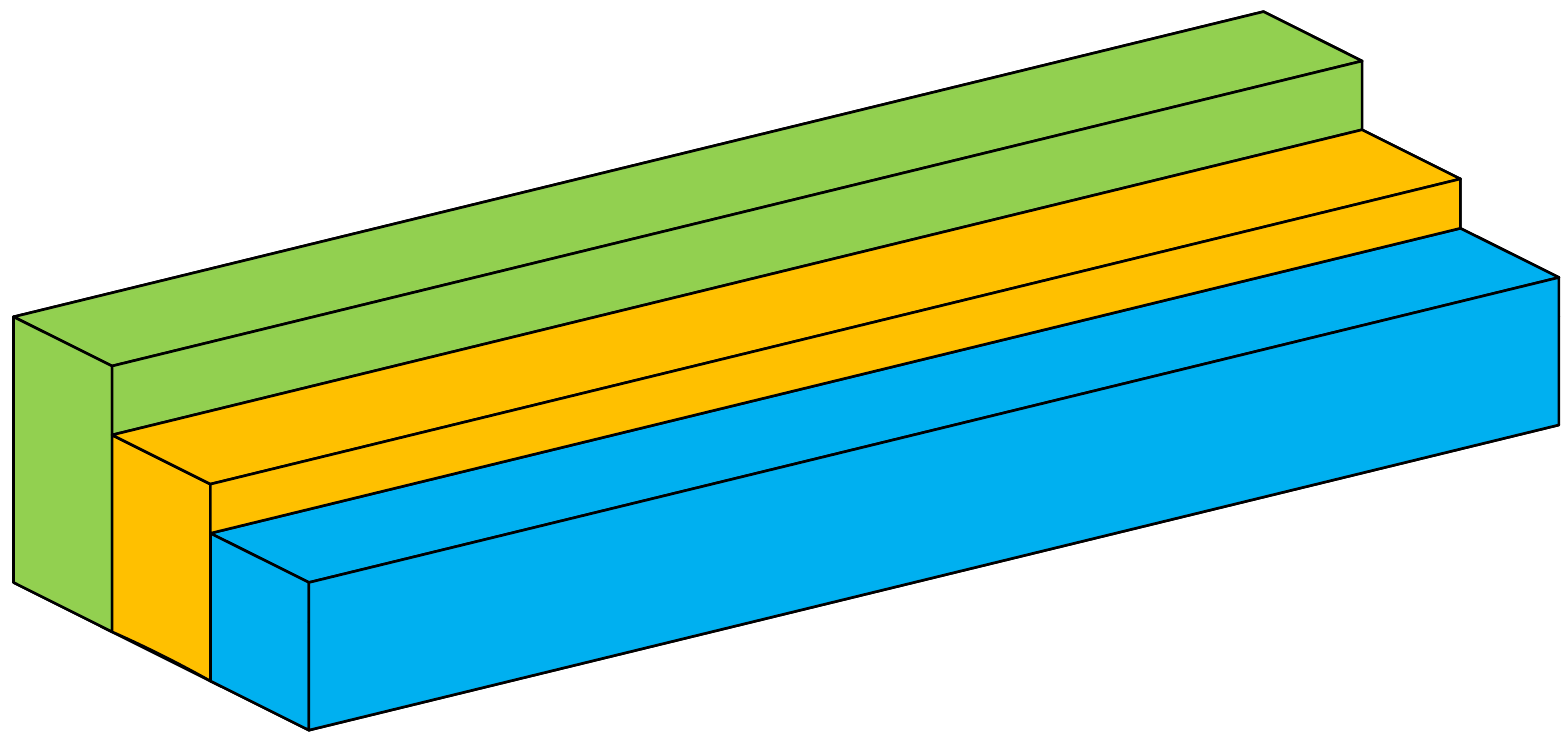}
				\put(50,18){$A$}
				\put(45,25){$B$}
				\put(40,32){$C$}
			\end{overpic}
		}
		\subfigure[]{
			\begin{overpic}[width=0.3\textwidth]{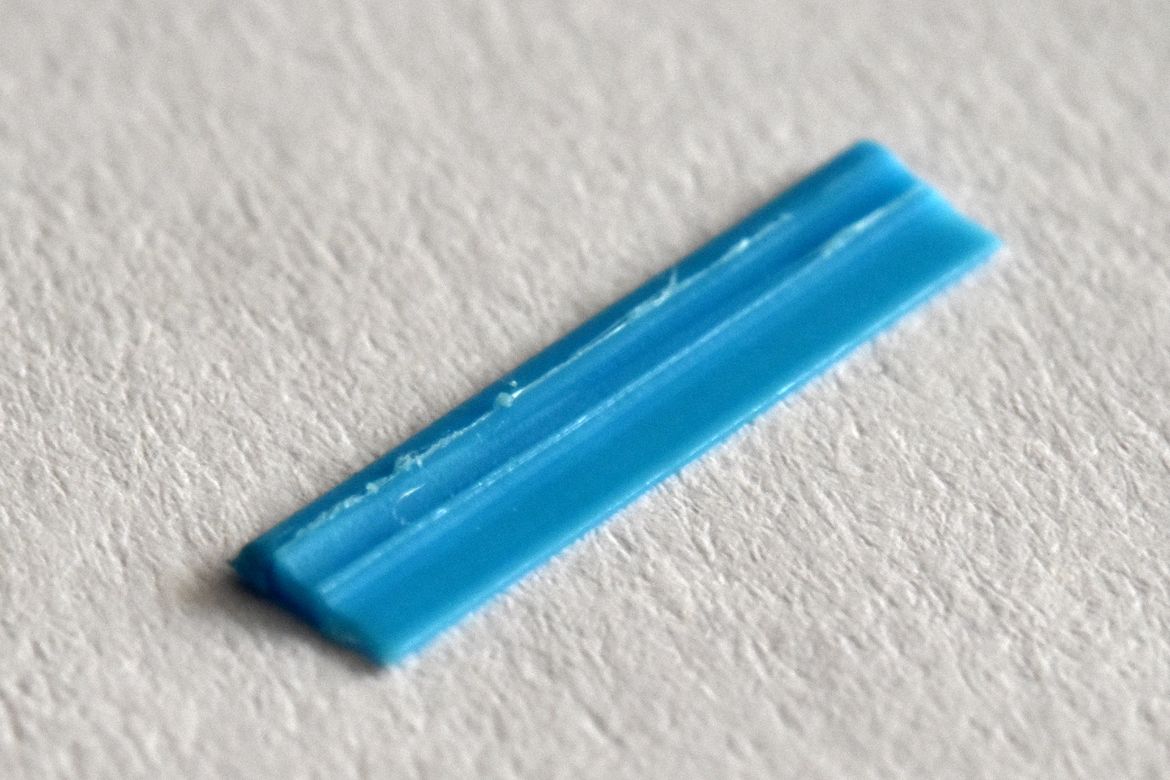}
			\end{overpic}
		}
		\subfigure[]{
			\begin{overpic}[width=0.3\textwidth]{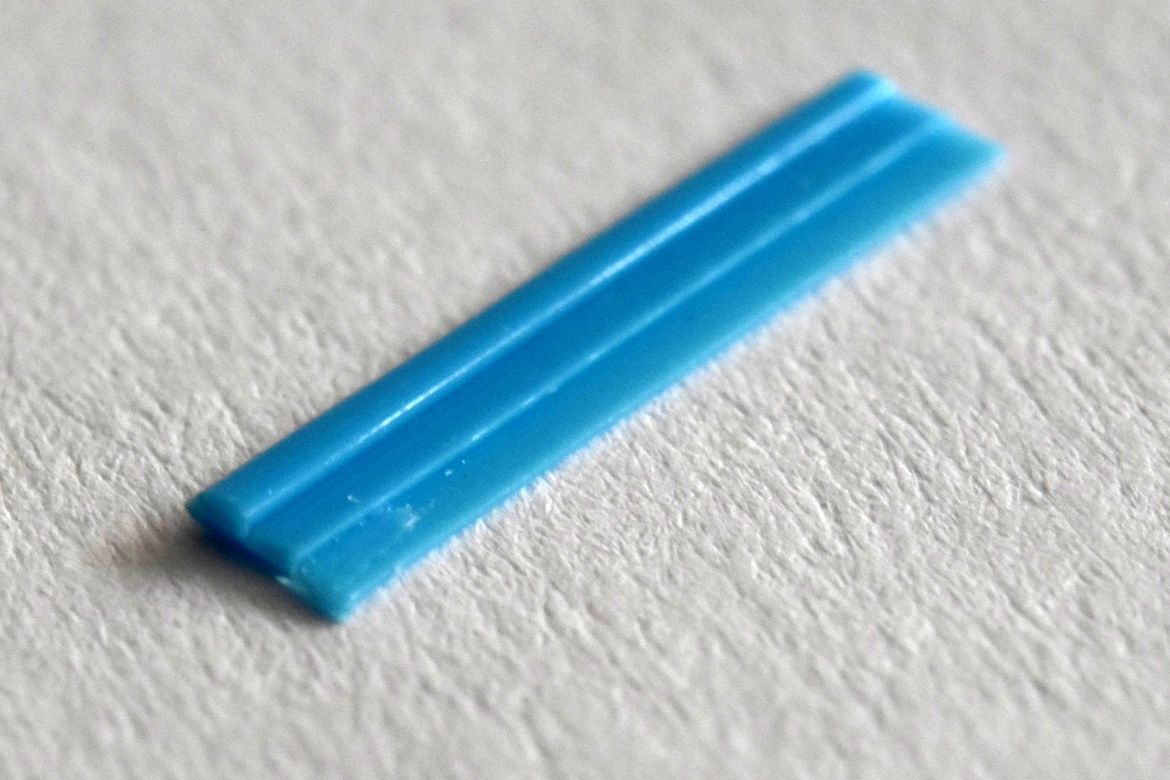}
			\end{overpic}
		}
		\caption{Interference between neighboring paths. \textbf{(a)} The heights of $A$, $B$ and $C$ are: 0.1 mm, 0.3 mm and 0.6 mm. The distances between the middle axis of $A$, $B$ and $C$ are $w=0.8$ mm, which is smaller than the nozzle outer diameter $\tau = 1.25$ mm. \textbf{(b)} Printed result in the order {$C$, $B$, $A$}. \textbf{(c)} Printed result in the order {$A$, $B$, $C$}}
		\label{fig:erase_path}
	\end{center}
\end{figure}
This has a detrimental effect on our approach, as the heated nozzle printing a low-height path can now interfere with a higher, previously printed path. This will damage the relief, effectively canceling
the slope that we are attempting to introduce.
This effect is visible in Figure~\ref{fig:erase_path}. Three neighboring paths $A$, $B$, $C$ are printed in two different orders. If higher paths are printed first, then the subsequently printed lower paths damage existing paths. However, if lower paths are printed first, then the subsequently printed higher paths will not interfere.

\begin{figure}\centering
	\begin{center}
		\begin{overpic}[width=0.35\textwidth]{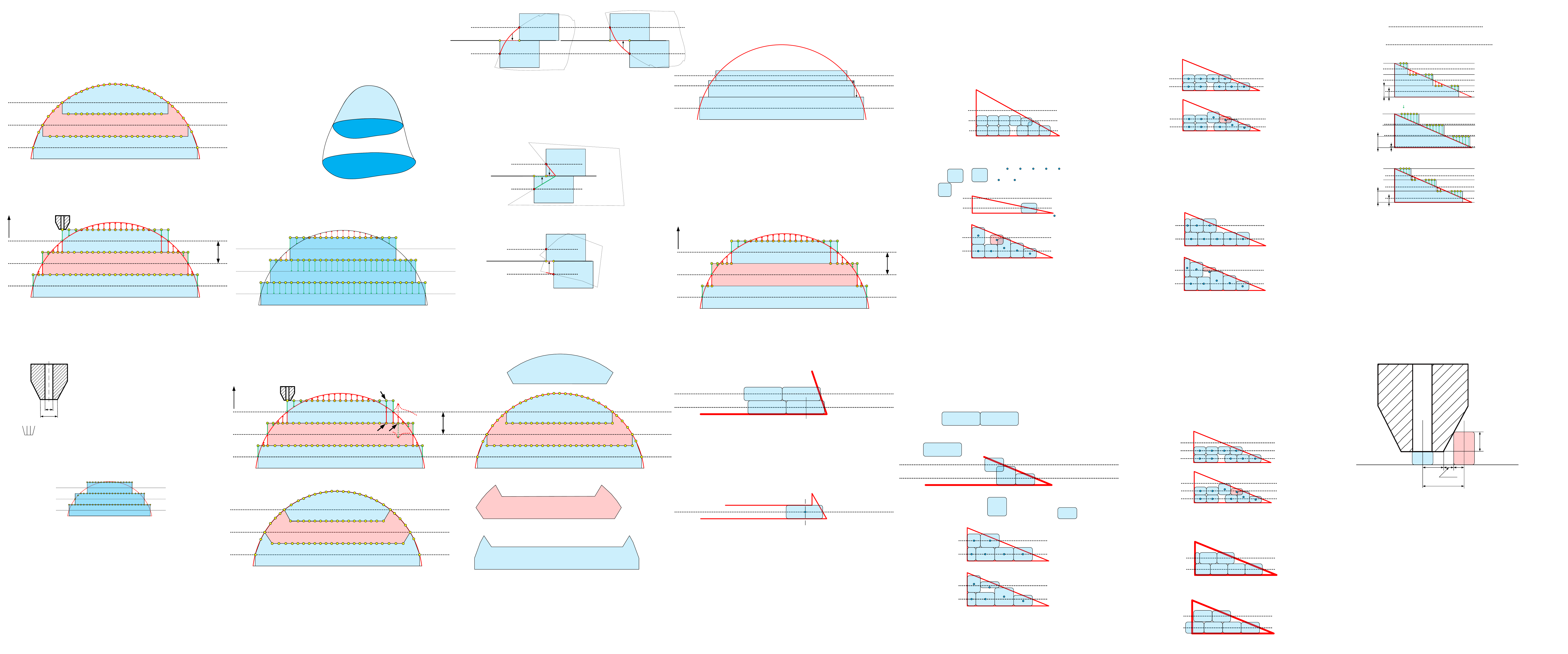}
			\put(74,17.5){\tiny$\cfrac{d}{2}$}
			\put(51,17.5){\tiny$\cfrac{\tau}{2}$}
            \put(60.5,14.5){\tiny$\Delta{h}\times{\cot}{\alpha}$}
            \put(48,7.5){\tiny$\epsilon={\frac{\tau+d}{2}+\Delta{h}\times{\cot}{\alpha}}$}
            \put(96,45){\tiny$\Delta{h}$}
		\end{overpic}
		\caption{Threshold computation for detecting interferences between paths (cross-section view). The blue rectangle is the cross section of a first path, the red rectangle is the cross section of a second path. $\tau$ is the outer diameter of the nozzle, $\alpha$ is the inclination of the nozzle side, $d$ is the path width. $\Delta{h}$ is the height difference between the two paths.}
	\label{fig:neighboring_threshold}
	\end{center}
\end{figure}


\begin{figure}\centering
	\begin{center}
		\subfigure[]{
			\begin{overpic}[width=0.4\textwidth]{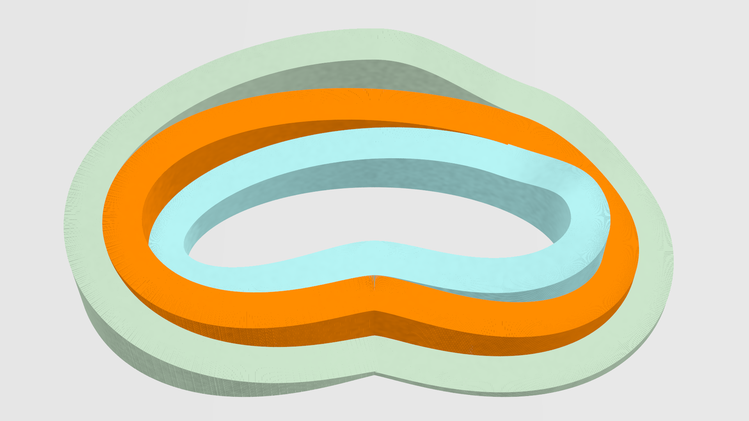}
				\put(40,48){$Path\ 1$}
				\put(42,41){$Path\ 2$}
				\put(45,34){$Path\ 3$}
			\end{overpic}
		}
		\subfigure[]{
			\begin{overpic}[width=0.4\textwidth]{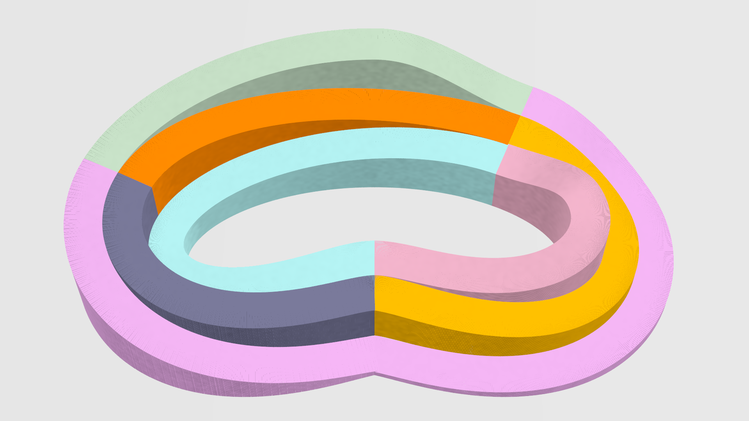}
				\put(50,7){$A$}
				\put(40,48){$B$}
				\put(32,13){$C$}
				\put(69,11){$D$}
				\put(42,41){$E$}
				\put(32,32){$F$}
				\put(76,25){$G$}
			\end{overpic}
		}
		\caption{Paths are split according to their relative heights. \textbf{(a)} Initial three paths. \textbf{(b)} Resulting seven subpaths after splitting.}
		\label{fig:split_path}
	\end{center}
\end{figure}
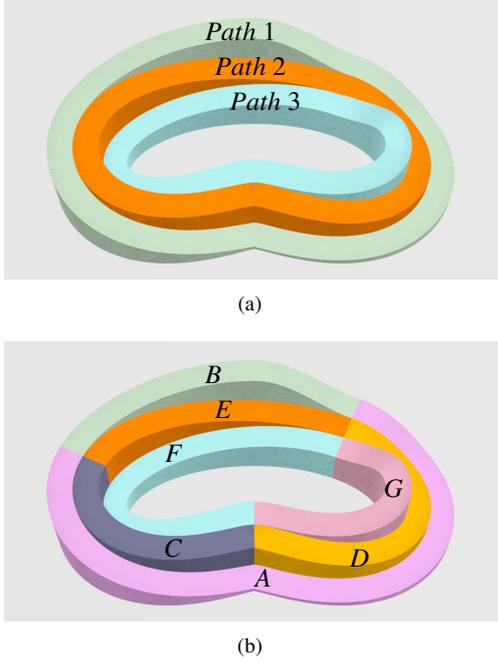

As just hinted, to print properly the potentially interfering paths have to be ordered from lower to higher heights. In particular, any two paths whose contours have a Hausdorff distance smaller than a threshold might interfere and have to be ordered. We compute the threshold as $\epsilon={\frac{\tau+d}{2}+\Delta{h}\cdot{\cot}{\alpha}}$
(this computation is illustrated in Figure \ref{fig:neighboring_threshold}). To obtain the final threshold we  set $\Delta{h}$ equal to the layer thickness, as our approach never displaces a tool path by more than the layer thickness. This gives a conservative, larger threshold that detects all possibly conflicting paths. The path width $d$ is generally determined by the inner nozzle width $w$, and thus we set $d=w$.

Each path can exhibit complex height variations, going up and down several times. Therefore, a path might be lower that its neighbor along a section and become higher later on. To make ordering possible we therefore split neighboring paths into subpaths that can be ordered. An example is shown in Figure \ref{fig:split_path}. In this Figure there are three curved paths: $Path\ 1$, $Path\ 2$, $Path\ 3$. $Path\ 1$ and $Path\ 2$ are direct neighbors, as well as $Path\ 2$ and $Path\ 3$. They will interfere with one another during printing. $Path\ 1$ and $Path\ 3$ are independent as they are too far from one another. As can be seen the paths go up and down several times, and therefore have to be split to allow for an ordering. $Path\ 1$ is split into $A$ and $B$, $Path\ 2$ is split into $C$, $D$ and $E$, and  $Path\ 3$ is split into $F$ and $G$.

Note that paths which were not modified by the anti-aliasing procedure do not need to be ordered and are printed first, in their initial ordering (even if they may be slightly damaged, these paths are hidden from view).


\subsection{Height constraint graph}

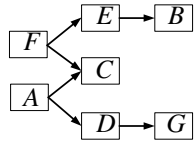
\begin{figure}\centering
    \begin{center}
      \begin{overpic}[width=0.15\textwidth]{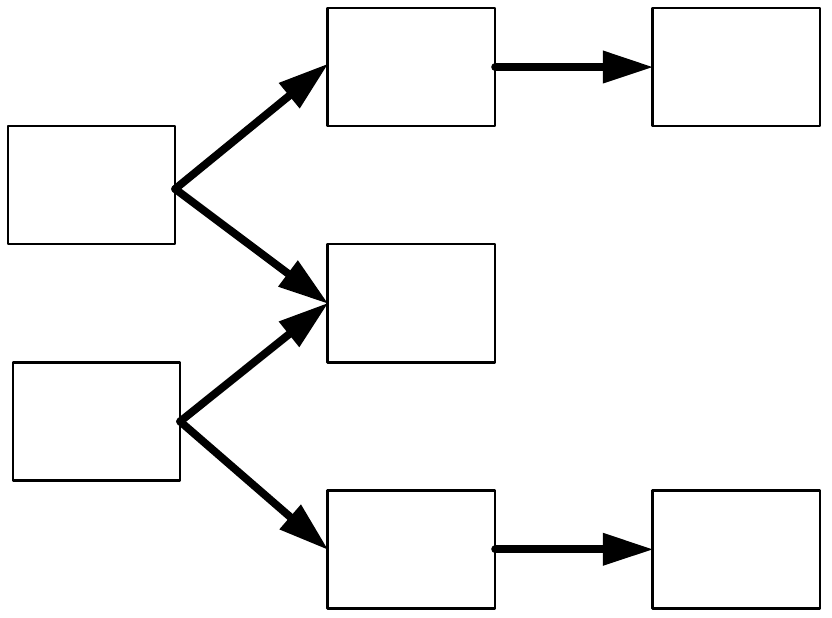}
       \put(8,19){$A$}
      \put(85,63){$B$}
      \put(47,5){$D$}
      \put(47,34){$C$}
      \put(47,63){$E$}
      \put(8,49){$F$}
      \put(85,5){$G$}
      \end{overpic}
\caption{Height constraint graph for the example in Figure~\ref{fig:split_path}}\label{fig:graph}
    \end{center}
\end{figure}

In order to search for an ordering we build a \textit{height constraint graph}. Each node of this directed graph corresponds to a printing path. Node $A$ has an edge pointing to node $B$ if the Hausdorff distance between path $A$ and path $B$ is smaller than $\tau$ \textit{and} path $A$ is lower than path $B$. The edge direction implies that $A$ has to be printed before $B$. The graph is shown in Figure \ref{fig:graph} for the example of Figure~\ref{fig:split_path}.

Paths are ordered by performing a topological sort on the height constraint graph. This is an iterative process that at each step selects a node with no edge pointing in ($0$ in-degree), adds it to the ordering and deletes all the edges going pointing out of it. The order in which we select the nodes is important for print quality and will be discussed next. This process always terminates as after splitting the graph is acyclic.

\subsection{Minimizing the number of seams}
\label{optimization}

Reducing interferences between paths is not the only concern to achieve a good surface quality. After printing each path the extruder has to travel to the next. If the start vertex $Q$ of the following path is far away from the end vertex $P$ of the current path, there will be small gaps left on $P$ and $Q$, respectively. These gaps are due to the interruption of the plastic extrusion, and produce visible seams on the final print. However, if $P$ and $Q$ approximately coincide, the transition will remain inconspicuous. Since we are splitting paths, the number of potential gaps increases. Therefore we aim to find an ordering sequence leaving a minimum number of gaps, from all possible topological sorts of the constraint height graph.


%

Let $N$ be the number of all path nodes in the height constraint graph. We use the following objective to represent the number of seams generated by a given path order $S=\{S_i|i=1,2,...,N\}$:

\begin{equation}
E(S)=\sum\limits_{i = 1}^{N-1} {NumGaps(S_i, S_{i+1}, G)}
\label{eqn:numgaps}
\end{equation}
where $NumGaps(S_i,S_{i+1},G)$ is the number of gaps generated from path $S_i$ to $S_{i+1}$ given the set of already produced gaps $G = Gaps(S_1,...,S_{i-1})$.
This is a set tracking the location of already introduced gaps. It lets us count gaps only once, giving the opportunity to exploit a previously introduced gap at no cost later in the sequence.

\paragraph*{Seam cost} $NumGaps(S_i,S_{i+1},G)$ evaluates to $0$ if the endpoint of path $S_i$ is near the entry point of $S_{i+1}$, as specified by a user specified tolerance $\epsilon$ (we use $\epsilon=4 \cdot w$).
Otherwise, it evaluates to the number of produced gaps -- we later describe an improvement where we weight the cost of the gaps, but for the sake of clarity we leave this out for now. Up to two gaps may appear (endpoint of $S_i$, entry point of $S_{i+1}$). These are only counted if they are not already in the set $G$ (e.g. they have not been already introduced earlier in the sequence). Therefore, $NumGaps$ returns a value in $\{0,1,2\}$.

\paragraph*{Searching for the best ordering}

The algorithm for searching the best ordering enumerates possible orders, tracking the one having minimal cost under Equation~\ref{eqn:numgaps}.
The number of orderings is limited since the only choices are between independent paths (e.g. A,F and E,C,D in Figure~\ref{fig:graph}) and only paths that have been touched by the anti-aliasing process are considered.
Nevertheless, to avoid a brute force search the algorithm prunes branches as early as possible. There are two pruning opportunities. The first occurs when the cost exceeds the current best. The second occurs when no two paths in the remaining set have matching exit/entry points. In such a case, there is no opportunity to factor gaps and the cost is the remaining number of exit/entry points (twice the number of remaining paths).

\begin{figure}[t]\centering
    \begin{center}
    \subfigure[]{
      \begin{overpic}[width=0.15\textwidth]{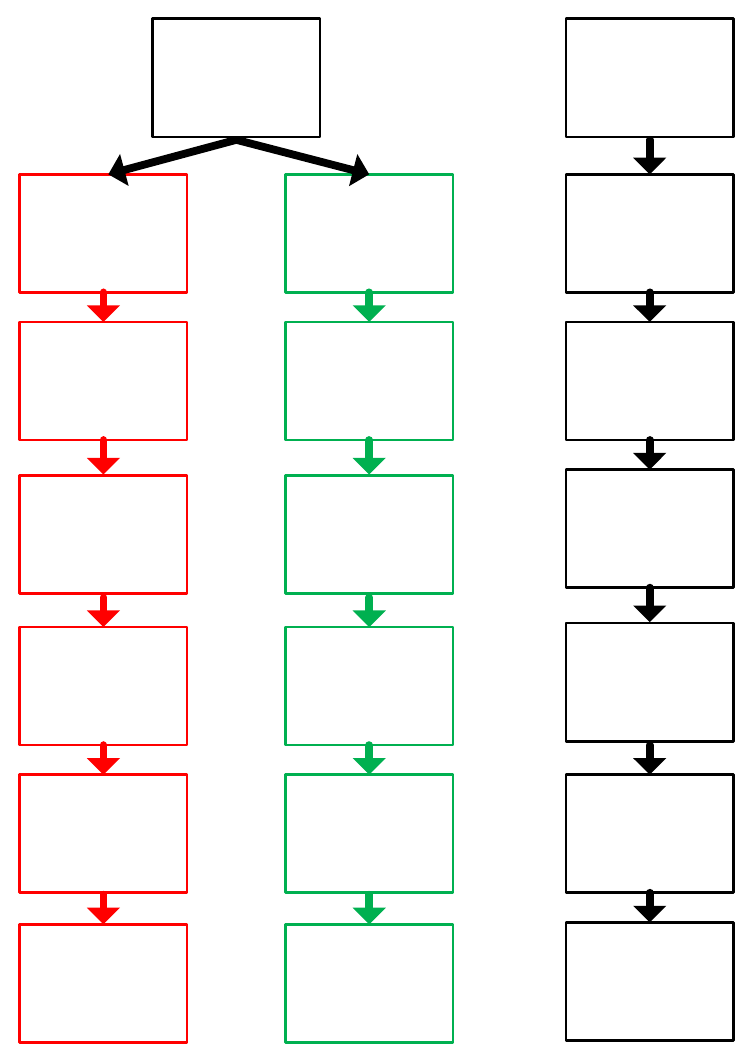}
       \put(18.5,89){$A$}
       \put(58,89){$F$}

       \put(6,74.4){$D$}
       \put(6,61){$F$}
       \put(6,46){$C$}
       \put(6,32){$E$}
       \put(6,18){$B$}
       \put(6,3.5){$G$}

       \put(31,74.4){$F$}
       \put(31,61){$D$}
       \put(31,46){$E$}
       \put(31,32){$B$}
       \put(31,18){$C$}
       \put(31,3.5){$G$}

       \put(57.5,74.4){$A$}
       \put(57.5,61){$E$}
       \put(57.5,46){$C$}
       \put(57.5,32){$D$}
       \put(57.5,18){$B$}
       \put(57.5,3.5){$G$}

       \put(6,85){$7$}
       \put(34,85){$3$}
       \put(58.5,100){$3$}

      \end{overpic}
      }
      \subfigure[]{
      \begin{overpic}[width=0.092\textwidth]{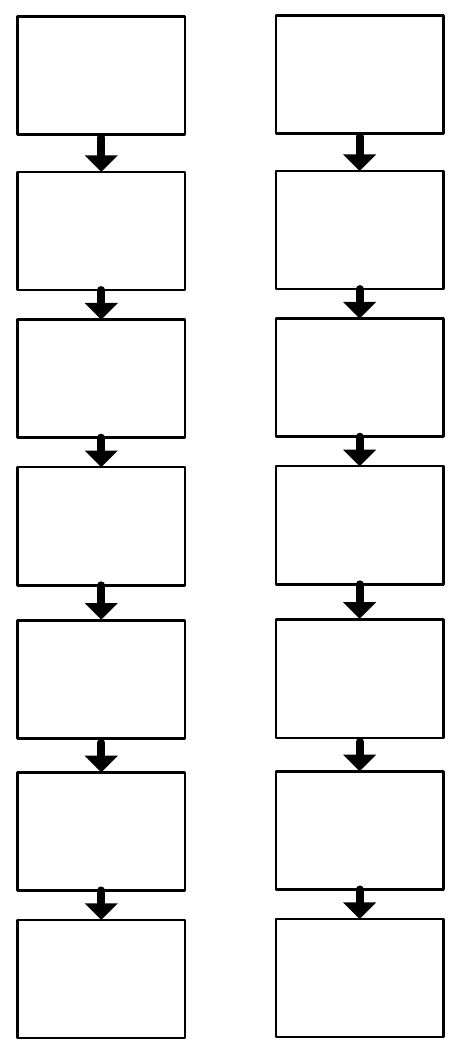}

       \put(6,89){$A$}
       \put(6,74.4){$F$}
       \put(6,61){$D$}
       \put(6,46){$E$}
       \put(6,32){$C$}
       \put(6,18){$G$}
       \put(6,3.5){$B$}

       \put(31,89){$F$}
       \put(31,74.4){$E$}
       \put(31,61){$A$}
       \put(31,46){$B$}
       \put(31,32){$C$}
       \put(31,18){$D$}
       \put(31,3.5){$G$}

       \put(4,100){$5.0$}
       \put(29.5,100){$4.5$}

      \end{overpic}
      }
		\vspace*{-4mm}
\caption{Ordering of paths in Figure \ref{fig:split_path}. \textbf{(a)} the order starting from $F$ and the green order starting from $A$ have the smallest number of gaps (3). The red order starting from $A$ has the largest (7). \textbf{(b)} When considering the concavity of the surface at the gaps, the order starting from $A$ has a score of $5.0$ while the order starting from $F$ has a smaller score of $4.5$.}\label{fig:seam_tree}
    \end{center}
\end{figure}
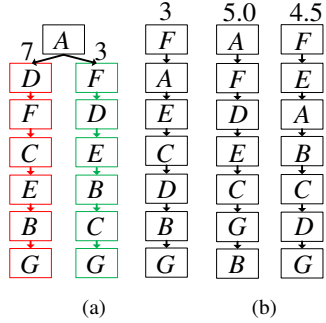

Examples are shown in Figure \ref{fig:seam_tree} (a) for the case of Figure~\ref{fig:split_path}. The path orders \{A,F,D,E,B,C,G\}, \{F,A,E,C,D,B,G\} both lead to the smallest number of gaps (3), while \{A,D,F,C,E,B,G\} gives the worst number of gaps (7). Let us denote $AB$ the end point of $A$ which is also the start point of $B$. The locations of the gaps generated by the two optimal path orders are respectively \{$AB$, $BA$, $GF$\} and \{$DE$, $FG$, $GF$\}. All other transitions do not introduce gaps.

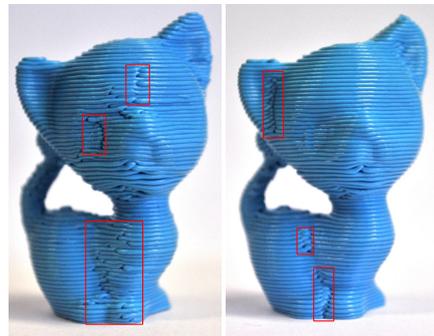
\begin{figure}[b]\centering
	\begin{center}
			\begin{overpic}[width=0.17\textwidth]{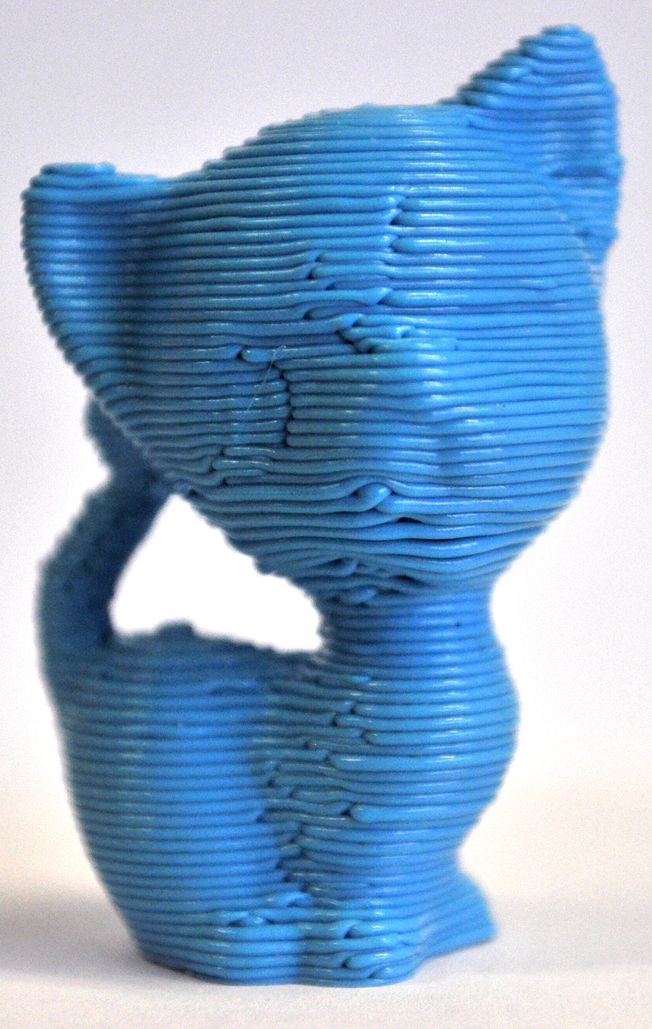}
				\put(35,70){\textcolor[rgb]{1.00,0.00,0.00}{\line(1,0){7}}}
				\put(42,70){\textcolor[rgb]{1.00,0.00,0.00}{\line(0,1){12}}}
				\put(42,82){\textcolor[rgb]{1.00,0.00,0.00}{\line(-1,0){7}}}
				\put(35,82){\textcolor[rgb]{1.00,0.00,0.00}{\line(0,-1){12}}}
				
				\put(22,55){\textcolor[rgb]{1.00,0.00,0.00}{\line(1,0){7}}}
				\put(29,55){\textcolor[rgb]{1.00,0.00,0.00}{\line(0,1){12}}}
				\put(29,67){\textcolor[rgb]{1.00,0.00,0.00}{\line(-1,0){7}}}
				\put(22,67){\textcolor[rgb]{1.00,0.00,0.00}{\line(0,-1){12}}}
				
				\put(23,4){\textcolor[rgb]{1.00,0.00,0.00}{\line(1,0){17}}}
				\put(40,4){\textcolor[rgb]{1.00,0.00,0.00}{\line(0,1){31}}}
				\put(40,35){\textcolor[rgb]{1.00,0.00,0.00}{\line(-1,0){17}}}
				\put(23,35){\textcolor[rgb]{1.00,0.00,0.00}{\line(0,-1){31}}}
			\end{overpic}
			\begin{overpic}[width=0.17\textwidth]{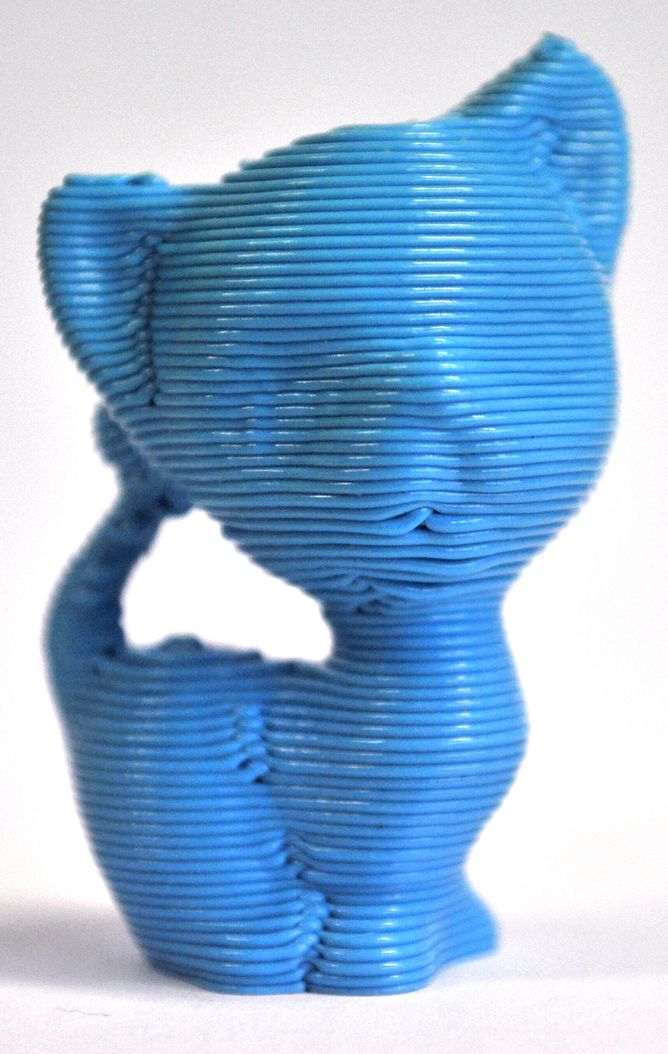}
				\put(11,60){\textcolor[rgb]{1.00,0.00,0.00}{\line(1,0){6}}}
				\put(17,60){\textcolor[rgb]{1.00,0.00,0.00}{\line(0,1){20}}}
				\put(17,80){\textcolor[rgb]{1.00,0.00,0.00}{\line(-1,0){6}}}
				\put(11,80){\textcolor[rgb]{1.00,0.00,0.00}{\line(0,-1){20}}}
				
				\put(21,25){\textcolor[rgb]{1.00,0.00,0.00}{\line(1,0){5}}}
				\put(26,25){\textcolor[rgb]{1.00,0.00,0.00}{\line(0,1){8}}}
				\put(26,33){\textcolor[rgb]{1.00,0.00,0.00}{\line(-1,0){5}}}
				\put(21,33){\textcolor[rgb]{1.00,0.00,0.00}{\line(0,-1){8}}}
				
				\put(26,5){\textcolor[rgb]{1.00,0.00,0.00}{\line(1,0){6}}}
				\put(32,5){\textcolor[rgb]{1.00,0.00,0.00}{\line(0,1){16}}}
				\put(32,21){\textcolor[rgb]{1.00,0.00,0.00}{\line(-1,0){6}}}
				\put(26,21){\textcolor[rgb]{1.00,0.00,0.00}{\line(0,-1){16}}}
			\end{overpic}
		\vspace*{-3mm}
		\caption{Hiding seams. Seams are highlighted with red rectangles. \textbf{Left:} No optimization of seams. \textbf{Right:} Result after our algorithm. Most of the visible
			gaps exist on the print without our technique (extrusion start/stops, also called zippers).}\label{fig:hide_seams}
	\end{center}
\end{figure}

\paragraph*{Hiding the gaps}
So far we have considered all gaps equivalent. However, some gaps have a stronger visual impact than others.
We draw inspiration from works on hiding start/end point of perimeter paths~\cite{hergel2014clean} to penalize gaps that are too visible.
In particular, gaps are easily noticed if positioned along smooth and convex regions, while they are less visible in concave regions of high curvature,
as shown in Figure \ref{fig:hide_seams}.

We therefore modify $NumGaps$ to return a sum of the 'cost' of each gap, which is defined as $(1 + \frac{\theta}{2\pi})$ with $\theta$ the angle opening to the outside of the model at
the gap location.
In Figure \ref{fig:seam_tree} (b), the orders \{A,F,D,E,C,G,B\} and \{F,E,A,B,C,D,G\} would normally both receive a cost of $5.0$. However, the second receives $4.5$ as the model is more concave at the produced gaps.

\section{Results}
\label{sec:results}


We evaluate our technique on a variety of examples. In all cases we use IceSL \cite{IceSL} to generate flat layer paths (but any other slicer could be used).
An Ultimaker 2 with PLA filament is used to print. It is fitted with a nozzle having an inner nozzle diameter $w=0.8$ mm, an outer nozzle diameter $\tau=1.25$ mm, and inclination $\alpha=45$ degrees. The base layer thickness is $0.6$ mm. All objects are printed with 100\% infilling. The z-motions are obtained by adding a z coordinate to all G1 motions in the G-code sent to the printer.

We perform all computations on an Intel Core i7, CPU 4.0 GHz, 32G memory.
All the slicing and printing times are recorded in Table \ref{time_comparison}.

We compare our algorithm with uniform slicing using flat layers at thicknesses $0.6$ mm and $0.2$ mm, which we denote respectively with labels flat06 and flat02 for compactness.

%
%

\paragraph*{Wedge}
This first example is shown in Figure~\ref{fig:teaser}. It reveals the improvement of layer anti-aliasing for a simple, elementary slope. Our result compares favorably to the print with $0.2$ mm layers (flat02) which takes \textit{twice} as long to print (using flat06 as reference, our print is only 6\% slower, while flat02 is 234 \% slower -- see Table~\ref{time_comparison}).
It takes 30 milliseconds for our anti-aliasing algorithm to process 327 vertices on 51 paths in 6 layers.

\begin{figure*}[h!]
  \centering
  \begin{tabular}{ccccc}
    \toprule
    Input model & Closeup view & flat06 & our algorithm & flat02 \\
     \midrule
     \moffsetmore{\multirow{4}{*}{\begin{overpic}[width=0.15\textwidth]{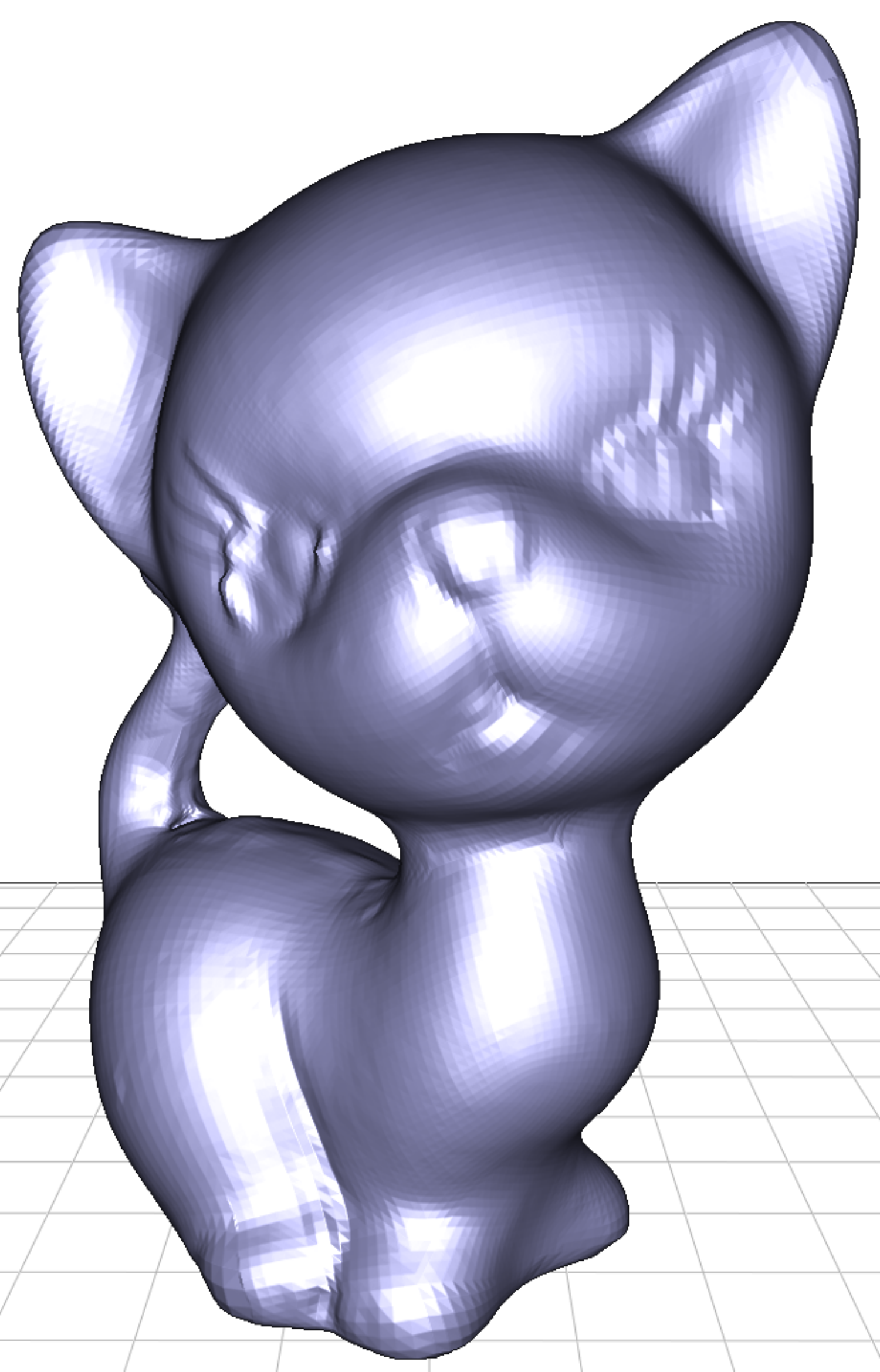}
      \put(25,87){\textcolor[rgb]{1.00,0.00,0.00}{\line(1,0){37}}}
      \put(62,87){\textcolor[rgb]{1.00,0.00,0.00}{\line(0,1){13}}}
      \put(62,100){\textcolor[rgb]{1.00,0.00,0.00}{\line(-1,0){37}}}
      \put(25,100){\textcolor[rgb]{1.00,0.00,0.00}{\line(0,-1){13}}}
      \put(12,36){\textcolor[rgb]{1.00,0.00,0.00}{\line(1,0){13}}}
      \put(25,36){\textcolor[rgb]{1.00,0.00,0.00}{\line(0,1){6}}}
      \put(25,42){\textcolor[rgb]{1.00,0.00,0.00}{\line(-1,0){13}}}
      \put(12,42){\textcolor[rgb]{1.00,0.00,0.00}{\line(0,-1){6}}}
      \put(62,87){\vector(4,-1){13}}
       \put(25,36){\vector(2,-1){49}}
      \end{overpic}}} &    \multirow{ 2}{*}{ \moffset{\begin{overpic}[width=0.15\textwidth]{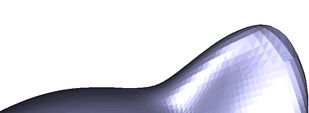} \end{overpic}}}
          & \mcenter{\centering\begin{overpic}[width=0.17\textwidth]{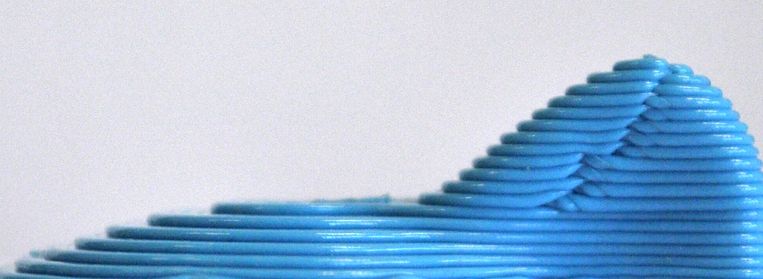} \put(10,35){\vector(1,-2){11}} \end{overpic} }& \mcenter{\centering\begin{overpic}[width=0.17\textwidth]{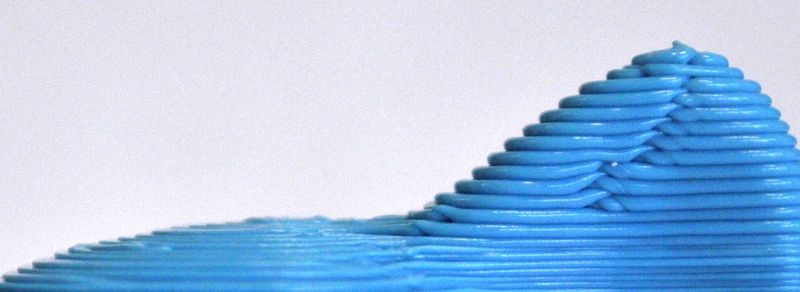} \put(10,35){\vector(1,-2){11}} \end{overpic} }& \mcenter{\centering\begin{overpic}[width=0.17\textwidth]{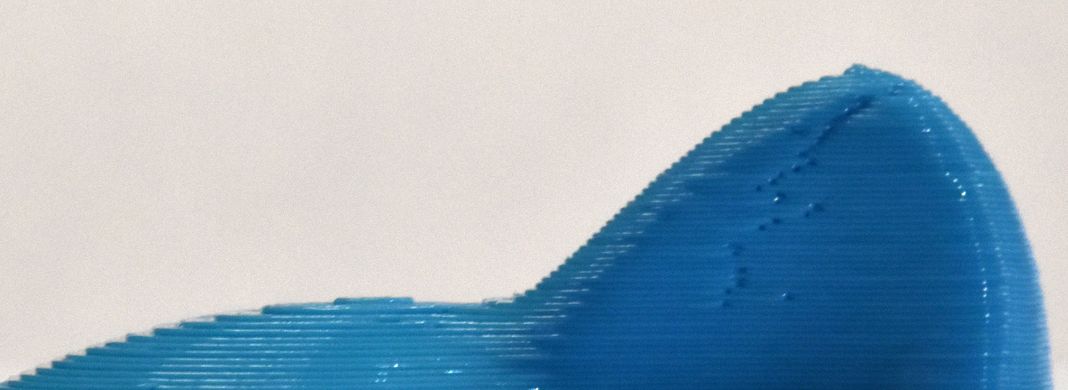} \put(10,35){\vector(1,-2){11}} \end{overpic} }     \\
     \mcenter{ }        & \mcenter{} & \mcenter{\centering\includegraphics[width=0.17\textwidth]{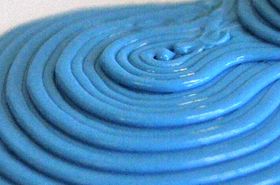}} & \mcenter{\centering\includegraphics[width=0.17\textwidth]{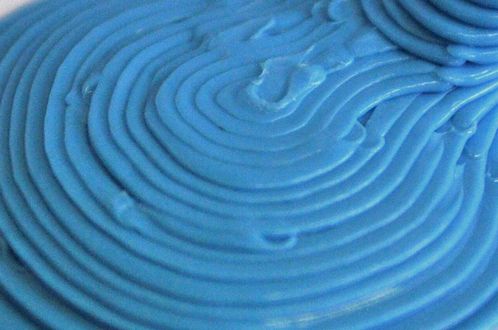} }& \mcenter{\centering\includegraphics[width=0.17\textwidth]{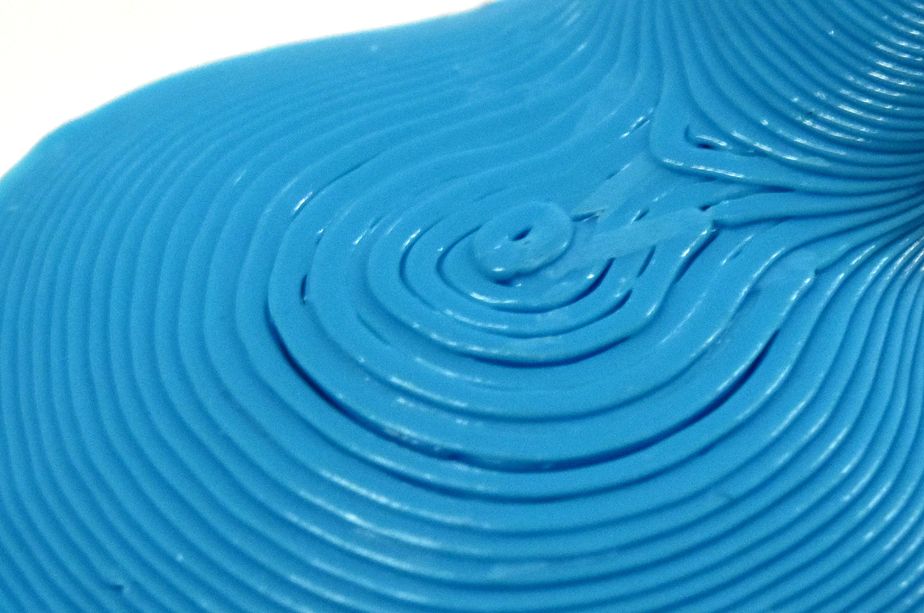}}\vspace{2pt}\\ \cline{2-5}
     &&&&\\
     \mcenter{} &\multirow{ 2}{*}{\moffset{\begin{overpic}[width=0.15\textwidth]{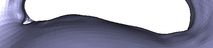} \end{overpic}}}
         & \mcenter{\centering\includegraphics[width=0.17\textwidth]{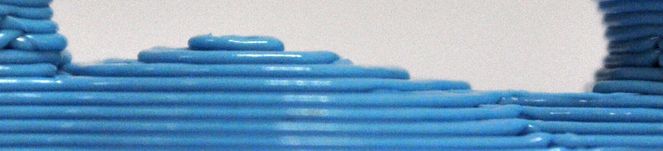} } & \mcenter{\centering\includegraphics[width=0.17\textwidth]{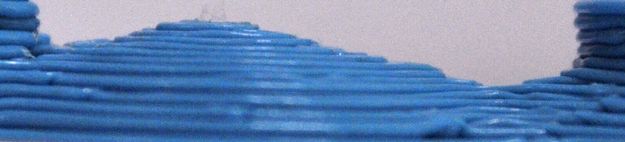} } & \mcenter{\centering\includegraphics[width=0.17\textwidth]{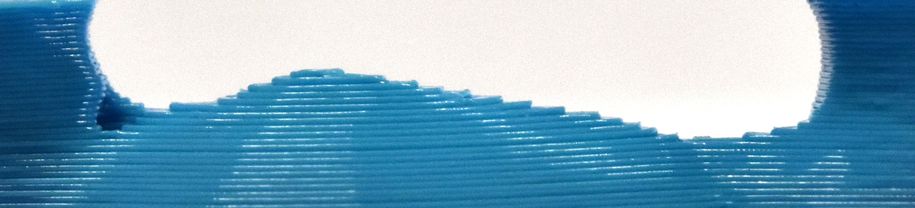}} \vspace{4pt}\\
     \mcenter{ }        & \mcenter{} & \mcenter{\centering\includegraphics[width=0.17\textwidth]{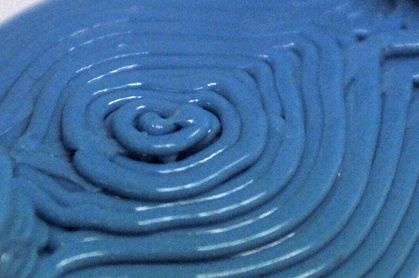} } & \mcenter{\centering\includegraphics[width=0.17\textwidth]{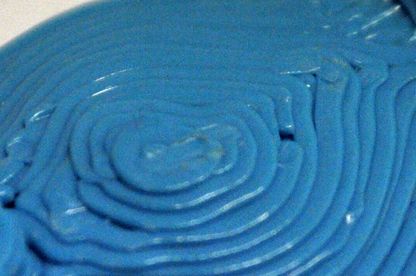} }& \mcenter{\centering\includegraphics[width=0.17\textwidth]{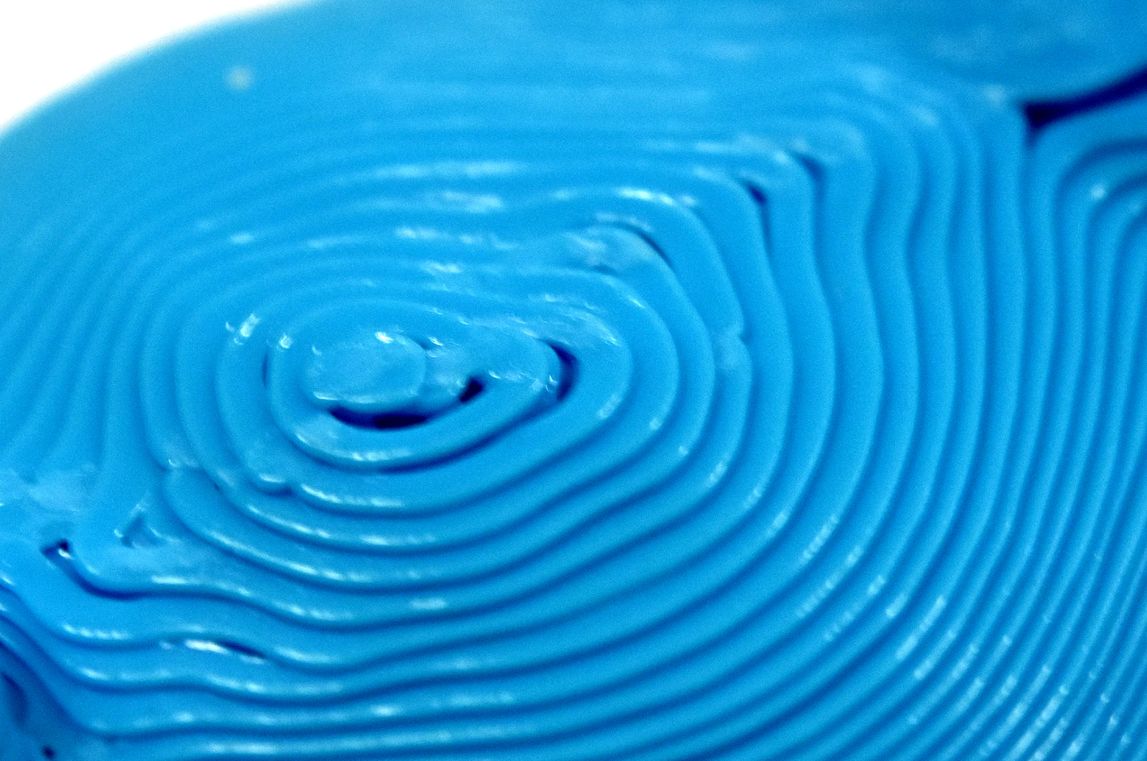}}\\
     \bottomrule
  \end{tabular}
  \caption{Real printouts of ``Kitten" (thing: 12694). Leftmost column: input 3D model and closeup regions outlined with red rectangles. Second column: 3D model, closeup views. Third column: flat06, side and top views on real printout. Fourth column: our result, side and top views. Fifth column: flat02, side and top views.}\label{fig:kitten_layer}
\end{figure*}

\paragraph*{Kitten}
Our second example is the ``Kitten" model, shown in Figure \ref{fig:kitten_layer}.
The print time of our result is only 2 \% slower than that of flat06, while flat02 is \textit{three times longer} to print.
Our result is more accurate than flat06 along silhouettes and gently slopped surfaces. Compared to flat02 silhouettes and gently slopped regions are either similar or slightly better: this is for instance visible in the top row of Figure \ref{fig:kitten_layer} (arrow).

The surface quality of our prints is affected by the additional gaps we introduce. However their number and impact is kept small as described in Section~\ref{optimization}. Figure~\ref{fig:hide_seams} shows the effect of our optimization on the result. Very careful calibration of the print parameters could also further reduce the gaps, but we generally assume this is too high a requirement from all but the most expert users.

It takes 314 milliseconds for our anti-aliasing algorithm to process 105466 vertices in 3078 paths and 146 layers.

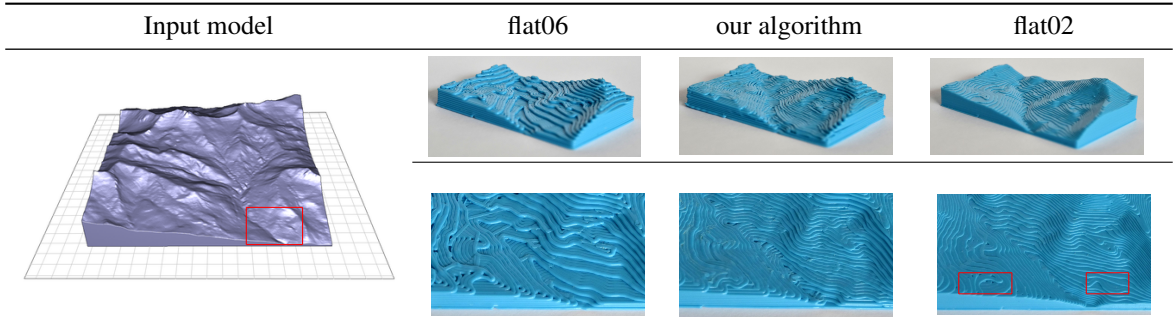
\begin{figure*}[h!]
  \centering
  \begin{tabular}{cccc}
    \toprule
    Input model & flat06 & our algorithm & flat02 \\
     \midrule
     \mcenter{\multirow{2}{*}{\begin{overpic}[width=0.3\textwidth]{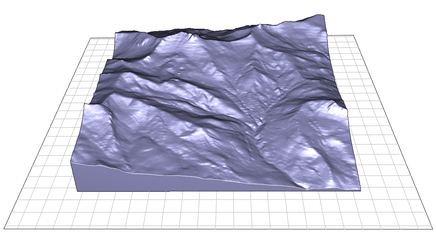}
      \put(60,10){\textcolor[rgb]{1.00,0.00,0.00}{\line(1,0){15}}}
      \put(75,10){\textcolor[rgb]{1.00,0.00,0.00}{\line(0,1){10}}}
      \put(75,20){\textcolor[rgb]{1.00,0.00,0.00}{\line(-1,0){15}}}
      \put(60,20){\textcolor[rgb]{1.00,0.00,0.00}{\line(0,-1){10}}}
      \end{overpic}}} &    \mcenter{\centering\begin{overpic}[width=0.17\textwidth]{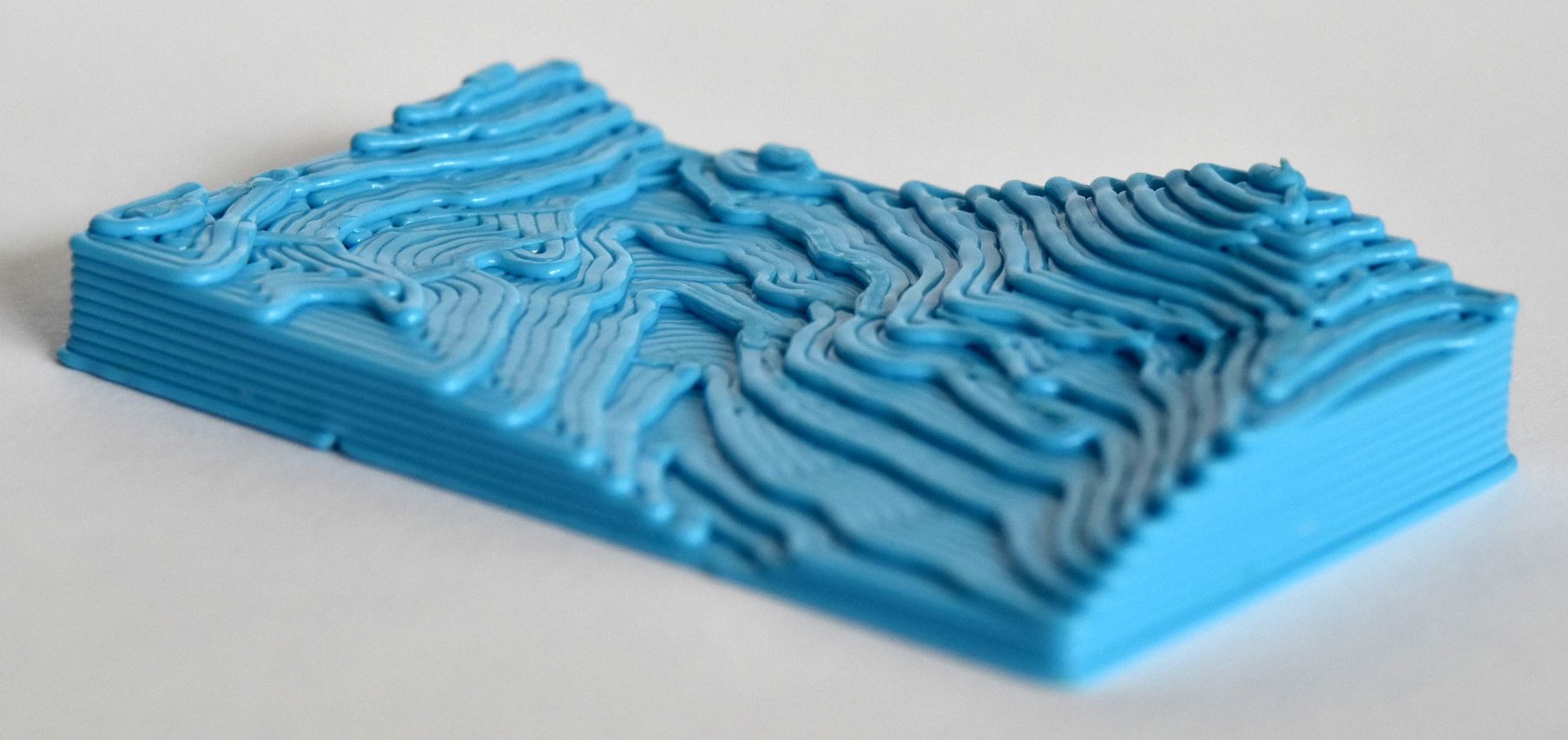} \end{overpic} }& \mcenter{\centering\begin{overpic}[width=0.17\textwidth]{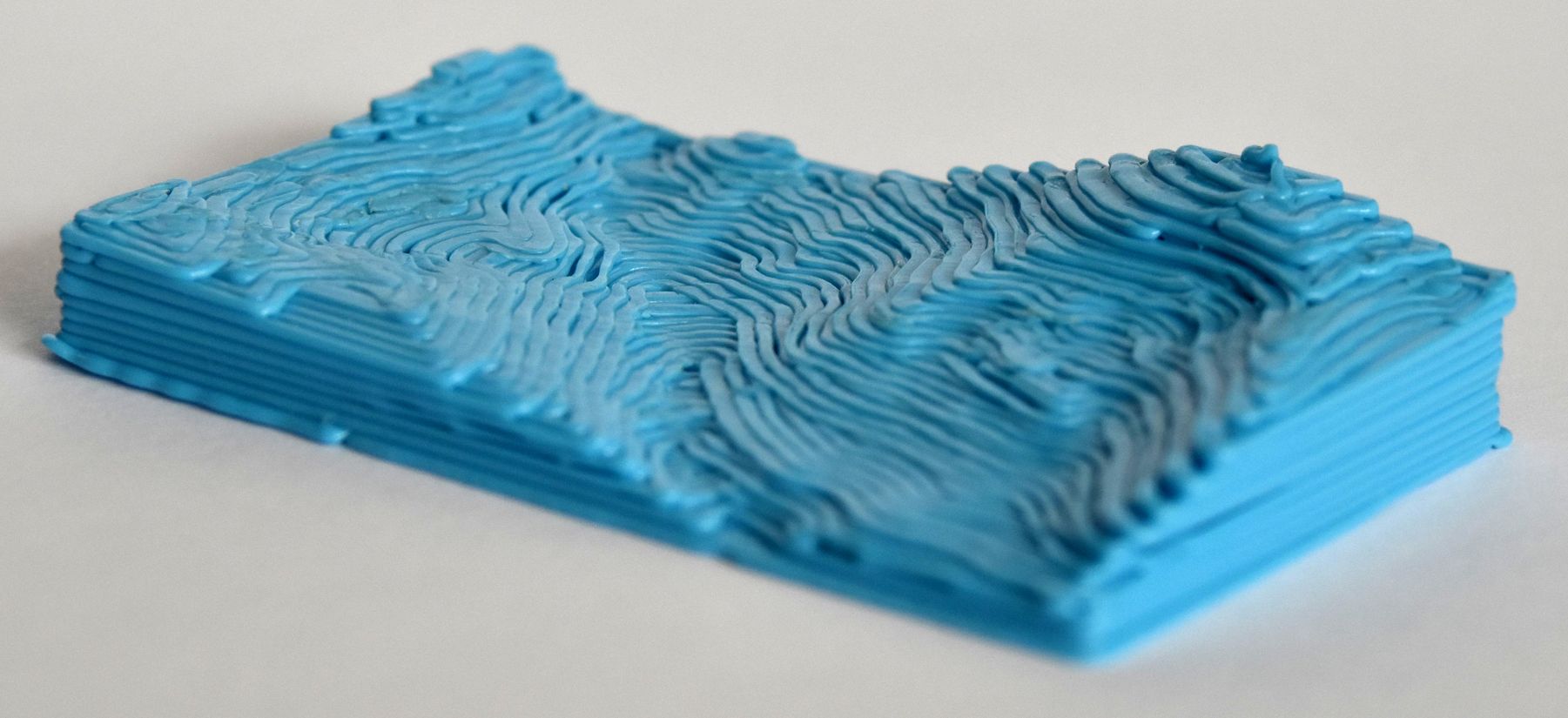} \end{overpic} }& \mcenter{\centering\begin{overpic}[width=0.17\textwidth]{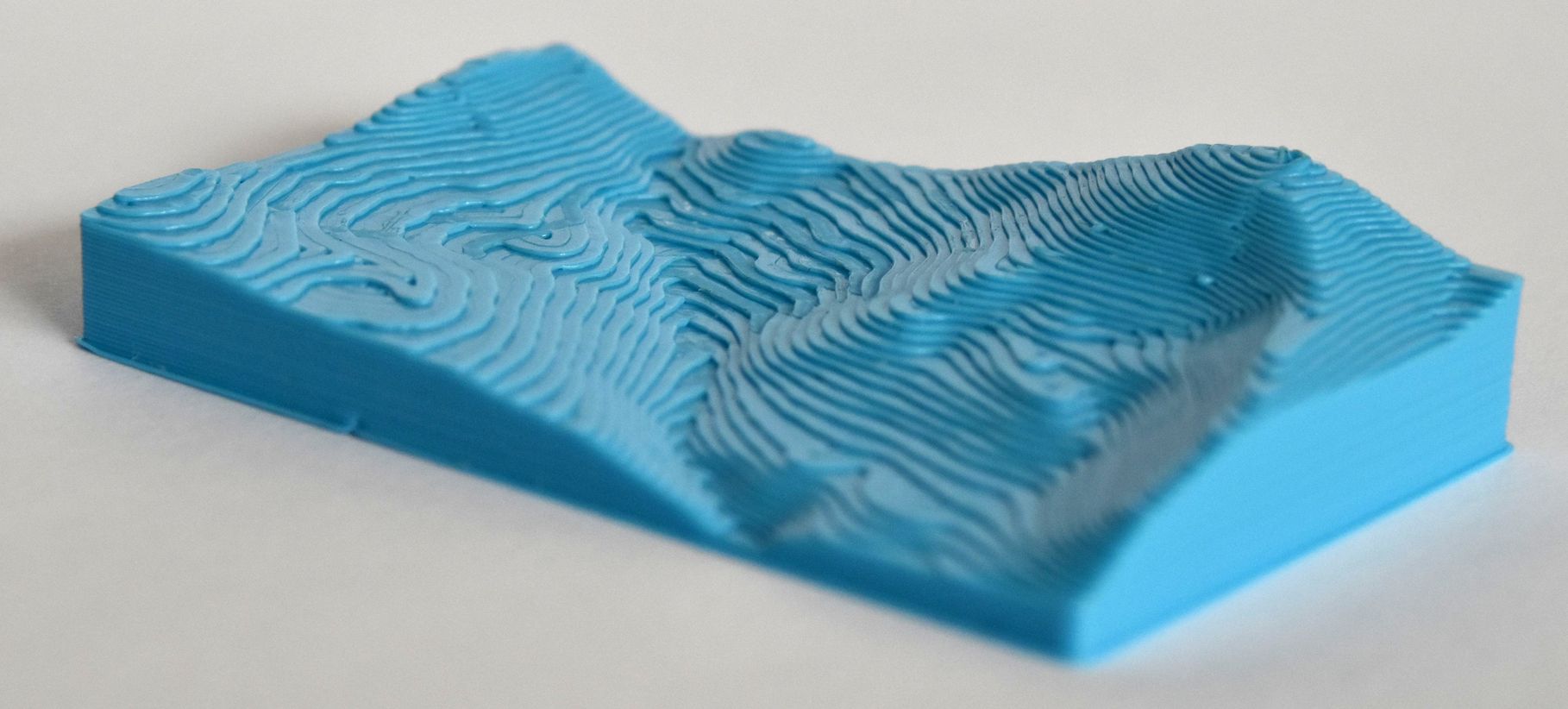} \end{overpic} }     \vspace{2pt}\\
      \cline{2-4}
      &&&\\
     \mcenter{ }  & \mcenter{\centering\includegraphics[width=0.17\textwidth]{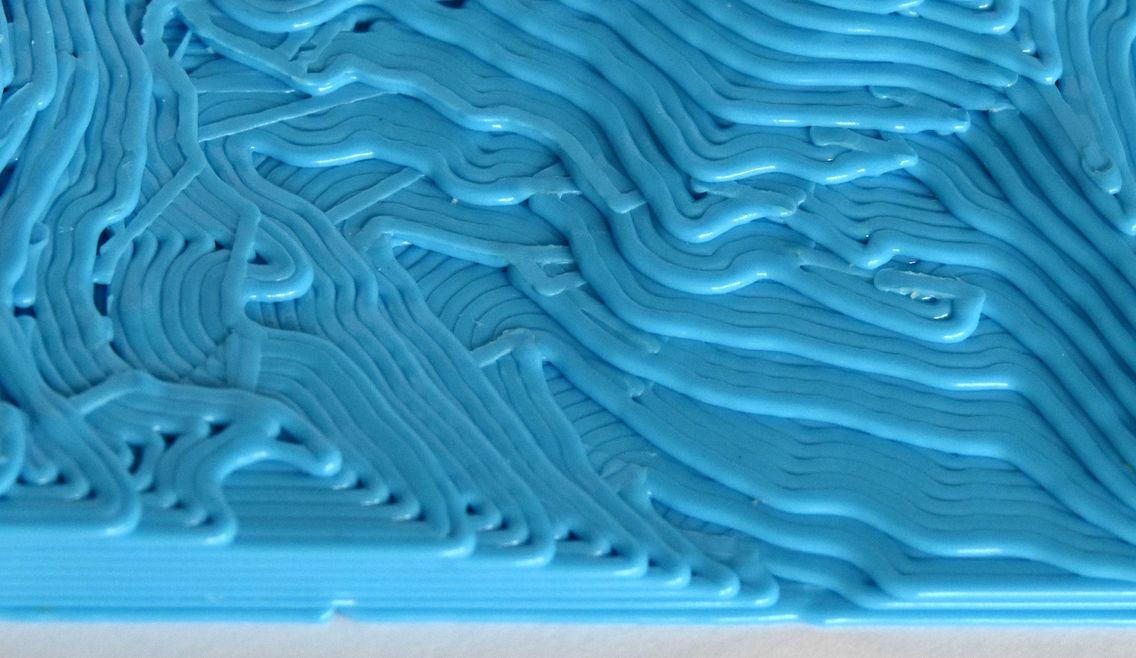}} & \mcenter{\centering\includegraphics[width=0.17\textwidth]{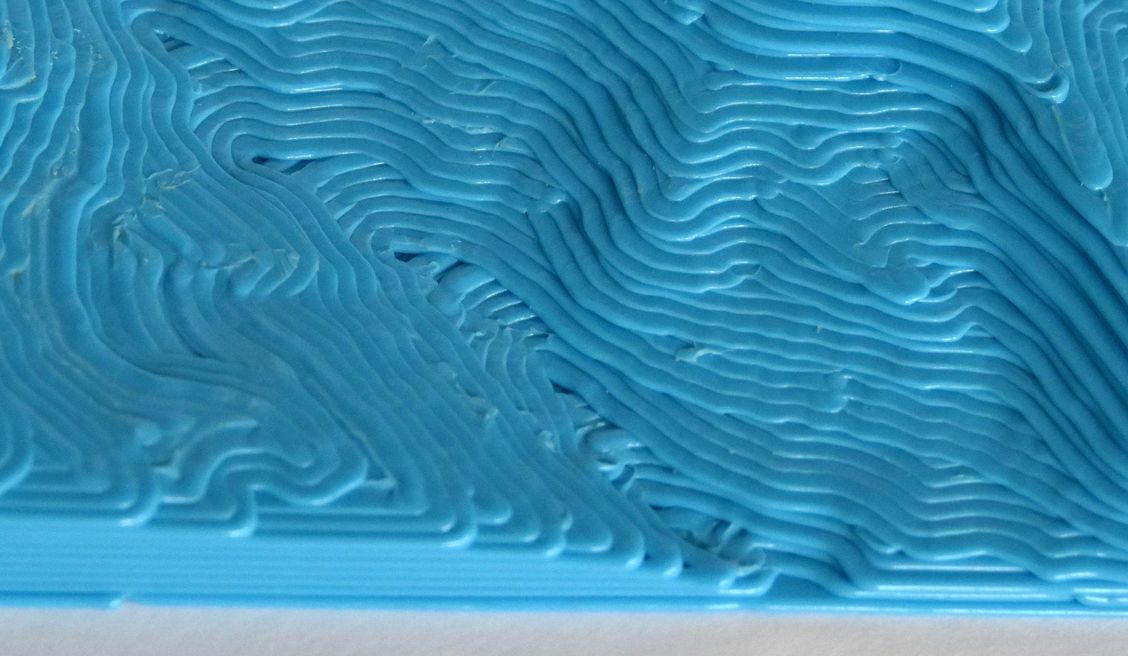} }& \mcenter{\begin{overpic}[width=0.17\textwidth]{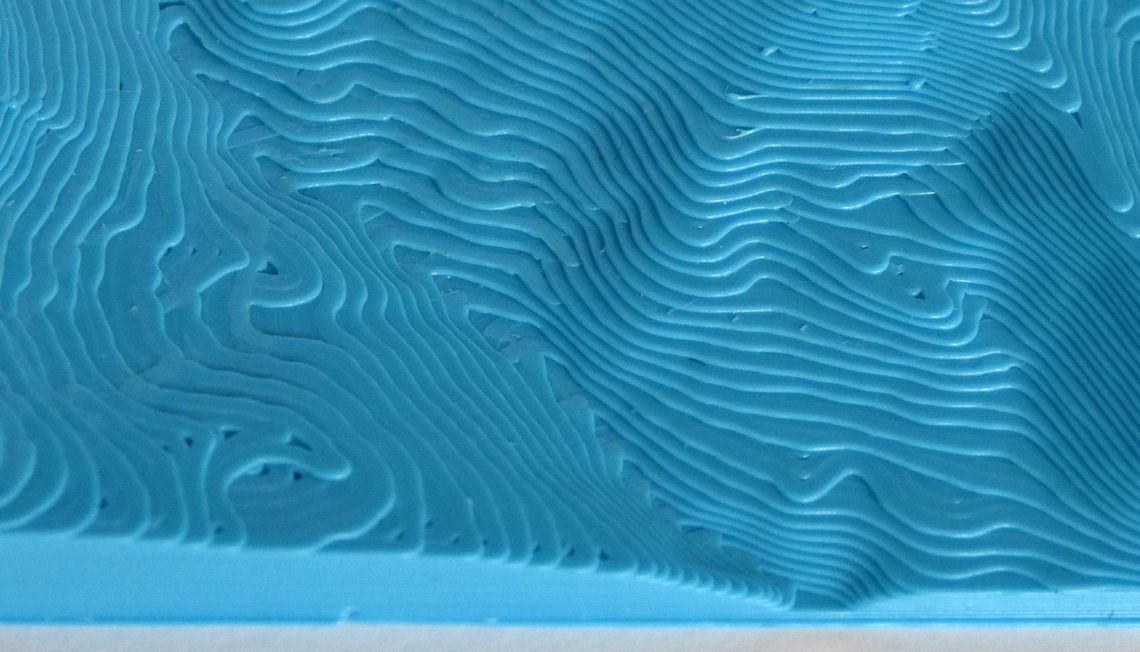}
      \put(70,10){\textcolor[rgb]{1.00,0.00,0.00}{\line(1,0){20}}}
      \put(90,10){\textcolor[rgb]{1.00,0.00,0.00}{\line(0,1){10}}}
      \put(90,20){\textcolor[rgb]{1.00,0.00,0.00}{\line(-1,0){20}}}
      \put(70,20){\textcolor[rgb]{1.00,0.00,0.00}{\line(0,-1){10}}}
      \put(10,10){\textcolor[rgb]{1.00,0.00,0.00}{\line(1,0){25}}}
      \put(35,10){\textcolor[rgb]{1.00,0.00,0.00}{\line(0,1){10}}}
      \put(35,20){\textcolor[rgb]{1.00,0.00,0.00}{\line(-1,0){25}}}
      \put(10,20){\textcolor[rgb]{1.00,0.00,0.00}{\line(0,-1){10}}}
      \end{overpic}}\vspace{2pt}\\
     \bottomrule
  \end{tabular}
  \caption{Real printouts of ``Lassen Volcanic" (thing: 667269). Leftmost column: input 3D model and closeup regions outlined with red rectangle. Second column: flat06, general and closeup views. Third column: our result, general and closeup views. Fourth column: flat06, general and closeup views.}\label{fig:lassen_result}
\end{figure*}

\paragraph*{Lassen Volcanic}
Our third example is a landscape with low slopes, as shown in Figure \ref{fig:lassen_result}.
We printed a subset of the whole model -- the red rectangle region in Figure \ref{fig:lassen_result}, left.
As shown in this figure, strong staircases exists in flat06. The surface quality is significantly improved in our result and is comparable with flat02.
It even slightly improves where staircases still exist in flat02, see regions highlighted by red rectangles in Figure \ref{fig:lassen_result}, right.
In this example, it takes 9.358 seconds for our anti-aliasing algorithm to process 211508 vertices on 10261 paths in 62 layers.

The print time of our result is nearly identical to that of flat06 (only 2\% slower), while flat02 takes more than twice as long to print (226\% slower than flat06).

\begin{figure*}[h!]
  \centering
  \begin{tabular}{cccc}
    \toprule
    Input model & flat06 & our algorithm & flat02 \\
     \midrule
     \mcenter{\multirow{2}{*}{\begin{overpic}[width=0.3\textwidth]{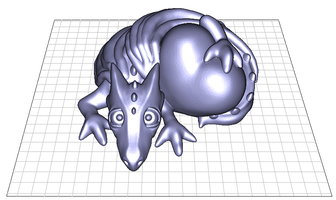}
      \put(30,27){\textcolor[rgb]{1.00,0.00,0.00}{\line(1,0){20}}}
      \put(50,27){\textcolor[rgb]{1.00,0.00,0.00}{\line(0,1){13}}}
      \put(50,40){\textcolor[rgb]{1.00,0.00,0.00}{\line(-1,0){20}}}
      \put(30,40){\textcolor[rgb]{1.00,0.00,0.00}{\line(0,-1){13}}}
      \end{overpic}}} &    \mcenter{\centering\begin{overpic}[width=0.17\textwidth]{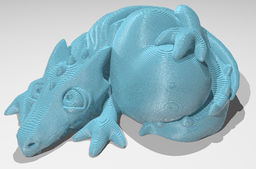} \end{overpic} }& \mcenter{\centering\begin{overpic}[width=0.17\textwidth]{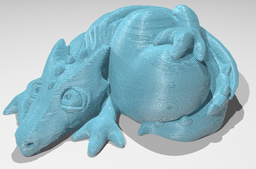} \end{overpic} }& \mcenter{\centering\begin{overpic}[width=0.17\textwidth]{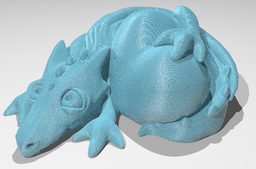} \end{overpic} }     \vspace{2pt}\\
      \cline{2-4}
      &&&\\
     \mcenter{ }  & \mcenter{\centering\includegraphics[width=0.17\textwidth]{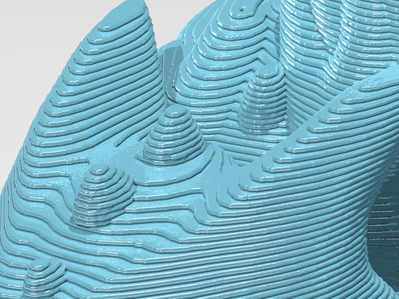}} & \mcenter{\centering\includegraphics[width=0.17\textwidth]{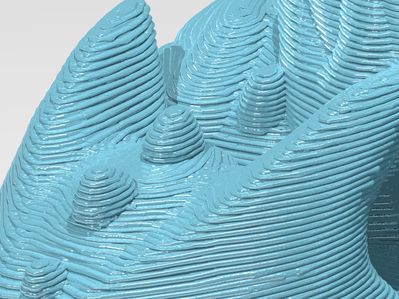} }& \mcenter{\begin{overpic}[width=0.17\textwidth]{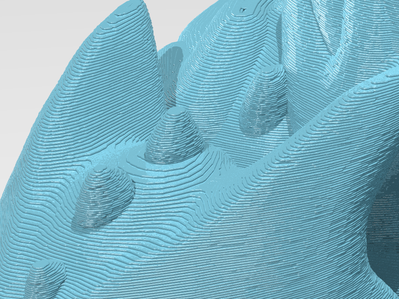}\end{overpic}}\vspace{2pt}\\
     \bottomrule
  \end{tabular}
  \caption{\textbf{Rendered} printouts of ``Pet monster Valentine" (thing: 17204). Leftmost column: input 3D model and closeup regions outlined with red rectangle. Second column: flat06, general and closeup views. Third column: our result, general and closeup views. Fourth column: flat02, general and closeup views.}\label{fig:pet}
\end{figure*}

\begin{figure*}[h!]
  \centering
  \begin{tabular}{cccc}
    \toprule
    Input model & flat06 & our algorithm & flat02 \\
     \midrule
     \mcenter{\multirow{2}{*}{\begin{overpic}[width=0.3\textwidth]{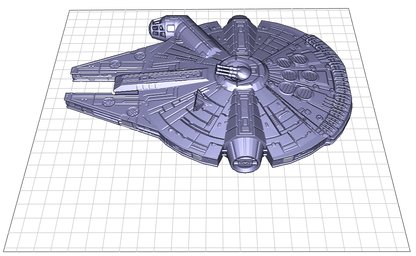}
      \put(45,22){\textcolor[rgb]{1.00,0.00,0.00}{\line(1,0){25}}}
      \put(70,22){\textcolor[rgb]{1.00,0.00,0.00}{\line(0,1){18}}}
      \put(70,40){\textcolor[rgb]{1.00,0.00,0.00}{\line(-1,0){25}}}
      \put(45,40){\textcolor[rgb]{1.00,0.00,0.00}{\line(0,-1){18}}}
      \end{overpic}}} &    \mcenter{\centering\begin{overpic}[width=0.17\textwidth]{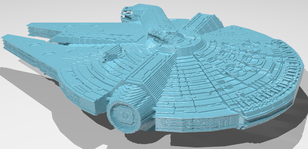} \end{overpic} }& \mcenter{\centering\begin{overpic}[width=0.17\textwidth]{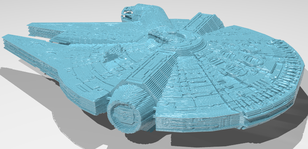} \end{overpic} }& \mcenter{\centering\begin{overpic}[width=0.17\textwidth]{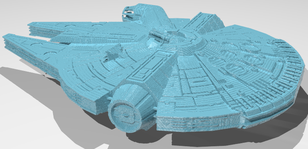} \end{overpic} }     \vspace{2pt}\\
      \cline{2-4}
      &&&\\
     \mcenter{ }  & \mcenter{\centering\includegraphics[width=0.17\textwidth]{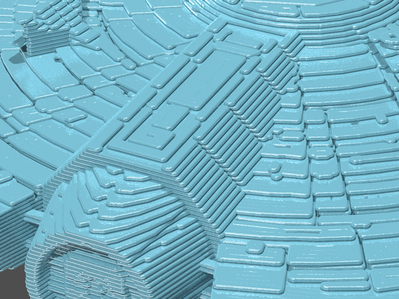}} & \mcenter{\centering\includegraphics[width=0.17\textwidth]{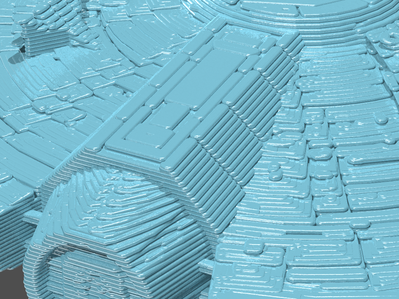} }& \mcenter{\begin{overpic}[width=0.17\textwidth]{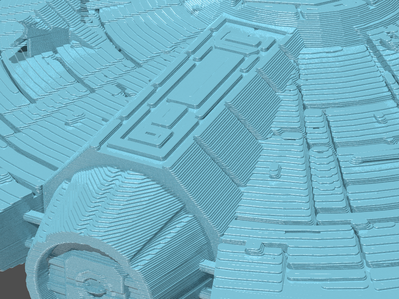}\end{overpic}}\vspace{2pt}\\
     \bottomrule
  \end{tabular}
  \caption{\textbf{Rendered} printouts of ``Fillenium Malcon" (thing: 919475). Leftmost column: input 3D model and closeup regions outlined with red rectangle. Second column: flat06, general and closeup views. Third column: our result, general and closeup views. Fourth column: flat02, general and closeup views.}\label{fig:fillenium}
\end{figure*}

\paragraph*{Rendered results}
We propose in Figures~\ref{fig:pet}, \ref{fig:fillenium} results which are rendered by our print simulator. This allows to appreciate the effect of the technique without
defects due to printer and filament calibration.

\paragraph*{Error maps}
To reveal the benefits in terms of surface error we compute error maps between the original surface and the surfaces produced by the toolpaths. The error maps are obtained by computing the distance from sampling points on the original model to the closest printed path, accounting for the deposited filament width and height. The distances range from $0.0$ mm to $0.3$ mm and are color coded.

\begin{figure*}\centering
    \begin{center}
    \subfigure[]{
      \begin{overpic}[width=0.25\textwidth]{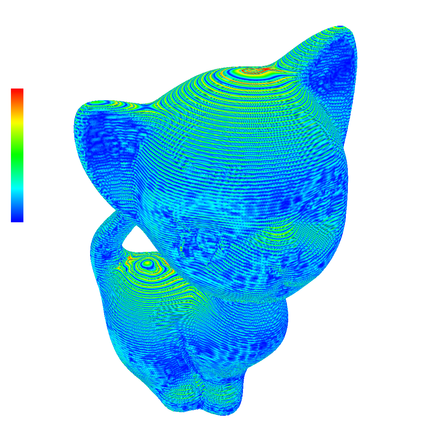}
      \put(-10,50){0.0}
      \put(-10,77){0.3}
      \end{overpic}
    }
    \subfigure[]{
      \begin{overpic}[width=0.25\textwidth]{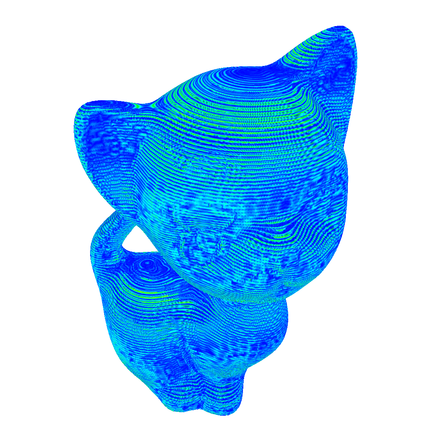}
      \end{overpic}
    }
    \subfigure[]{
      \begin{overpic}[width=0.25\textwidth]{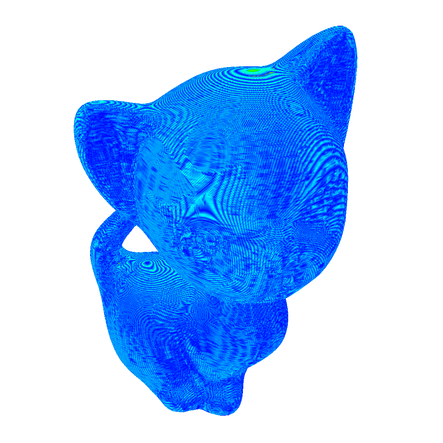}
      \end{overpic}
    }

    \subfigure[]{
      \begin{overpic}[width=0.25\textwidth]{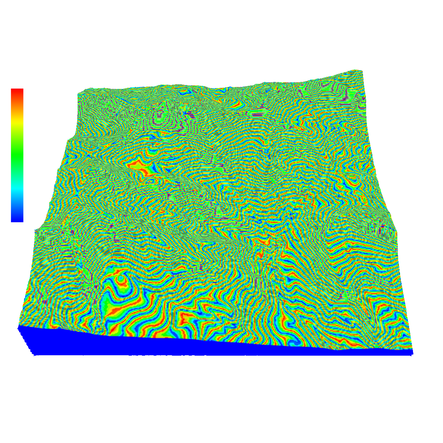}
      \put(-10,50){0.0}
      \put(-10,77){0.3}
      \end{overpic}
    }
    \subfigure[]{
      \begin{overpic}[width=0.25\textwidth]{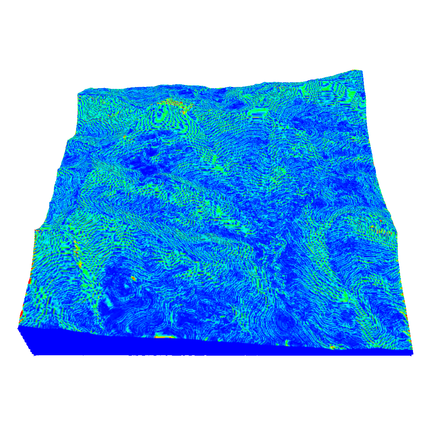}
      \end{overpic}
    }
    \subfigure[]{
      \begin{overpic}[width=0.25\textwidth]{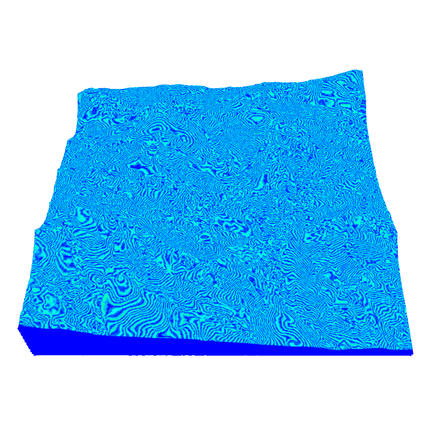}
      \end{overpic}
    }

    \subfigure[]{
      \begin{overpic}[width=0.25\textwidth]{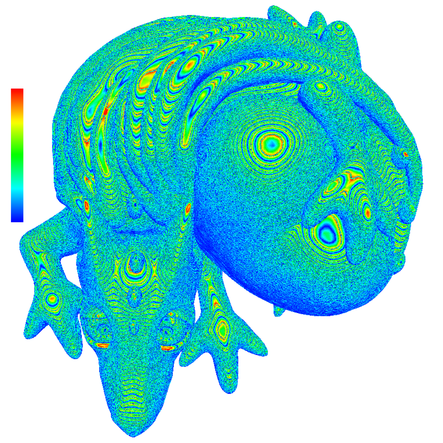}
      \put(-10,50){0.0}
      \put(-10,77){0.3}
      \end{overpic}
    }
    \subfigure[]{
      \begin{overpic}[width=0.25\textwidth]{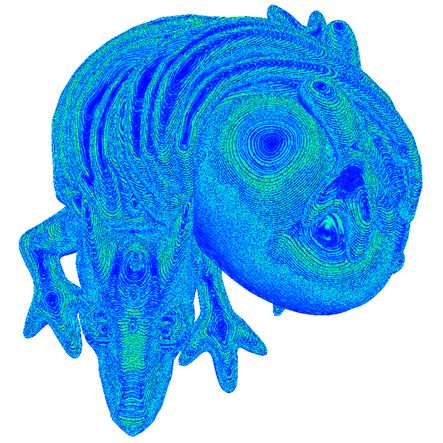}
      \end{overpic}
    }
    \subfigure[]{
      \begin{overpic}[width=0.25\textwidth]{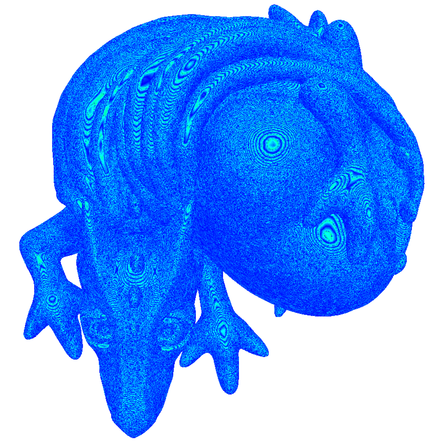}
      \end{overpic}
    }

    \subfigure[]{
      \begin{overpic}[width=0.25\textwidth]{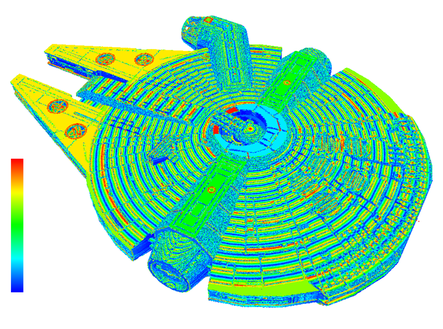}
      \put(-10,8.5){0.0}
      \put(-10,35.5){0.3}
      \end{overpic}
    }
    \subfigure[]{
      \begin{overpic}[width=0.25\textwidth]{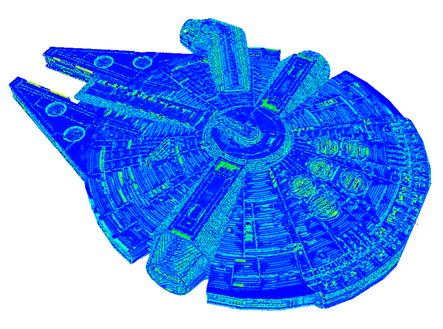}
      \end{overpic}
    }
    \subfigure[]{
      \begin{overpic}[width=0.25\textwidth]{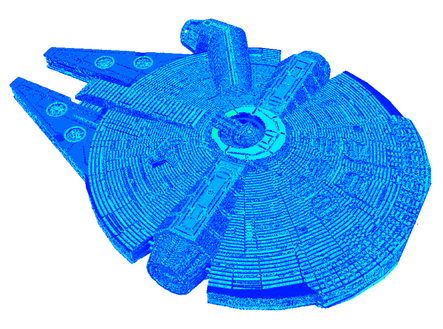}
      \end{overpic}
    }
    \caption{Error map of ``Kitten": (a) flat06, (b) our algorithm, (c) flat02; ``Lassen Volcanic": (d) flat06, (e) our algorithm, (f) flat02; ``Pet monster Valentine": (g) flat06, (h) our algorithm, (i) flat02; ``Fillenium Malcon": (j) flat06, (k) our algorithm, (l) flat02}\label{fig:error_kitten}
    \end{center}
\end{figure*}

Figure~\ref{fig:error_kitten} compares the error maps of flat06, our technique and flat02 for the ``Kitten", ``Lassen Volcanic", ``Pet monster Valentine", and ``Fillenium Malcon" models. As can be seen our technique greatly reduces the error along gently curved surfaces compared to flat06 (e.g. the top and back of ``Kitten", the plateaus of ``Lassen Volcanic"). These areas are also improved when compared to flat02, even though regions at greater angles still have more error (this is also visible on the printouts, e.g. the kitten ear in Figure~\ref{fig:kitten_layer} top row).

\begin{figure} 
\begin{center}
      \begin{overpic}[width=0.2\textwidth]{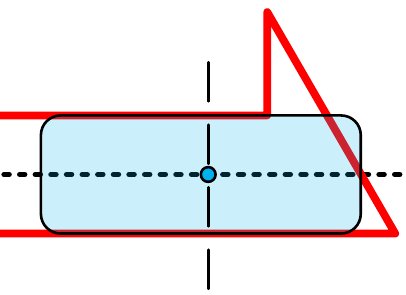}
      \end{overpic}
\caption{The blue rectangle is a cross-section of a printed plastic path. The red outline is the shape of the anti-aliased layer. The only degree of freedom to deform the path is the blue vertex. In such a case, the toolpaths cannot be modified to reproduce the sharp features of the anti-aliased layers.}
\label{fig:limitation}
\end{center}
\end{figure}

The error map of our result for ``Kitten" (denoted (b) in Figure~\ref{fig:error_kitten}) reveals a limitation of our technique. The top of the model has a greatly reduced error, but the error increases again when the angle grows beyond a threshold. This is due to the fact that the deposited plastic tracks are too large in these regions to reproduce the sharp features of the anti-aliased layer relief, as illustrated in Figure~\ref{fig:limitation} and discussed in Section~\ref{sec:layer2toolpaths}.

Generally the distance between neighboring paths is determined by the inner nozzle diameter $w$. Let the original layer thickness to be $h$. For an arbitrary vertex on the model surface, the angle between the tangent plane and the XY plane should be at most $\theta = \arctan(\frac{h}{w})$ to make infill paths available for anti-aliasing. With our setup this gives a critical angle of $37$ degrees ($w=0.8, h=0.6$).

Nevertheless, our technique greatly reduces the error where it is worst in flat06, while preserving a nearly identical print time.

%
%

\begin{table*}\footnotesize
     \caption{\label{time_comparison}Slicing time (in seconds) / printing time (in seconds) / G-Code file size (in MBs) comparisons between: flat06, flat02 and our algorithm. \#Order is the number of complete orders considered by our algorithm.}
     \begin{center}
     \begin{tabular}{ccccc}
      \toprule
      Cases & flat06 & flat02 & our algorithm & \#Order\\
      \midrule
      Wedge & 0.625~/~61 & 1.927~/~143 & 0.625+0.03~/~64&38\\
      Kitten & 57.772~/~4767 & 156.690~/~14234 & 57.772+0.314~/~4859&40293\\
      Lassen Volcanic & 342.474~/~25083 & 1039.691~/~66731 & 342.474+9.358~/~30434&256219\\
      Pet monster Valentine & 421.455~/~35637 & 1418.160~/~105470 & 421.455+25.815~/~36107&164494\\
      Fillenium Malcon & 183.858~/~15602 & 573.254~/~46860 & 183.858+25.515~/~15614&56363\\
      \bottomrule
     \end{tabular}
     \end{center}
\end{table*}

\section{Conclusion}

We proposed a technique to reduce the staircase aliasing defects that plague filament deposition. Our approach proposes to exploit small, sub-layer z-motions to better reproduce gently sloped areas. It does not require to globally curve the layers, nor does it require specific changes to existing printers. In particular our technique considers within-layer nozzle interference, splitting and reordering paths to minimize its detrimental effect.

The main limitation of our technique stems from the discretization of the layers into contours (plastic tracks) of a same width, as discussed in Section~\ref{sec:layer2toolpaths} and Section~\ref{sec:results}.
A second drawback of our approach is that it introduces additional gaps onto the print surface, even though their number of location is optimized to reduce their impact. Overall, the improvement in accuracy and visual quality of the silhouettes is very significant. 

Finally, it is worth noting that our technique can be applied on the output of any slicer, as long as both the G-Code and the original mesh are available. Therefore, we hope to see a wide adoption of this approach, which produces more accurate and visually pleasing surfaces at negligible cost in print time.

\section*{Acknowledgments}
This work was funded by ERC ShapeForge StG-2012-307877 with support from the R\'{e}gion Lorraine.

\section*{References}
\bibliographystyle{elsarticle-num}
\bibliography{biblio}

\end{document}